\documentclass[12pt]{article}
\usepackage[english]{babel}
\usepackage[T1]{fontenc}
\usepackage{graphicx}
\usepackage{amssymb,amsthm,amsmath,amsfonts}
\usepackage{wasysym}
\usepackage{harvard}
\usepackage{ae}   
\usepackage{aecompl} 
\usepackage{float}
\usepackage{setspace}

\newcommand{\ud}{\mathrm{d}}
\newcommand{\vect}[1]{\boldsymbol{\mathrm{#1}}}

\renewcommand{\deg}{^\circ}

\newcommand{\AU}{\,\mathrm{AU}}
\newcommand{\km}{\,\mathrm{km}}
\newcommand{\meter}{\,\mathrm{m}}
\newcommand{\cm}{\,\mathrm{cm}}

\newcommand{\yr}{\,\mathrm{yr}}
\newcommand{\days}{\,\mathrm{d}}

\newcommand{\second}{\,\mathrm{s}}

\newcommand{\rev}{\,\mathrm{rev}}

\begin{document}


\citationstyle{agsm}
\citationmode{abbr}

\mbox{ }
\vspace{2cm}
\begin{center}
{\bf{\Large The population of natural Earth satellites}} \\

\vspace{0.5cm}
Mikael Granvik${}^\mathrm{a,b}$, Jeremie Vaubaillon${}^\mathrm{c}$, Robert Jedicke${}^\mathrm{a}$ \\
\vspace{0.5cm}
E-mail: mgranvik@iki.fi \\
\vspace{0.5cm}
${}^\mathrm{a}$Institute for Astronomy, University of Hawaii, 2680 Woodlawn Drive, Honolulu, HI 96822, U.S.A. \\
${}^\mathrm{b}$Department of Physics, P.O. Box 64, 00014 University of Helsinki, Finland \\
${}^\mathrm{c}$Institut de M\'ecanique C\'eleste et de Calcul des \'Eph\'em\'erides, Observatoire de Paris, 77 
Avenue Denfert-Rochereau, F-75014 Paris, France
\\
\vspace{0.5cm}
Submitted to \emph{Icarus} on August 19, 2011 \\
Accepted for publication on December 13, 2011 \\
\vspace{1cm}
Manuscript pages: 63 \\
Tables:  3 \\
Figures: 29 \\
\vspace{1cm}
Version: \today \\
\end{center}

\newpage

\mbox{ }
\vspace{0cm}
\begin{center}
{\bf Proposed Running Head:} The population of natural Earth satellites \\
\end{center}
\vspace{10cm}
{\bf Editorial correspondence to:} \\
Mikael Granvik \\
Department of Physics \\
P.O. Box 64 \\
00014 University of Helsinki \\
Finland \\
Phone: +358 (0)9 191 50751 \\
Fax: +358 (0)9 191 50610 \\
E-mail: mgranvik@iki.fi

\newpage

\begin{abstract}
\noindent We have for the first time calculated the population
characteristics of the Earth's irregular natural satellites (NES) that
are temporarily captured from the near-Earth-object (NEO) population.
The steady-state NES size-frequency and residence-time distributions
were determined under the dynamical influence of all the massive
bodies in the solar system (but mainly the Sun, Earth, and Moon) for
NEOs of negligible mass. To this end, we compute the NES capture
probability from the NEO population as a function of the latter's
heliocentric orbital elements and combine those results with the
current best estimates for the NEO size-frequency and orbital
distribution. At any given time there should be at least one NES of
1-meter diameter orbiting the Earth. The average temporarily-captured
orbiter (TCO; an object that makes at least one revolution around the
Earth in a co-rotating coordinate system) completes
$(2.88\pm0.82)\rev$ around the Earth during a capture event that lasts
$(286\pm18)\days$. We find a small preference for capture events
starting in either January or July. Our results are consistent with
the single known natural TCO, 2006 RH$_{120}$, a few-meter diameter
object that was captured for about a year starting in June 2006. We
estimate that about 0.1\% of all meteors impacting the Earth were
TCOs.
\end{abstract}
\newpage

\section{Introduction}
\label{sec:introduction}

In this work we provide the first study and characterization of the
population of temporarily-captured natural Earth satellites (NES;
Table \ref{table:acronyms} contains a list of all acronyms used in the
paper) including their steady-state size-frequency distribution (SFD),
capture probability as a function of the near-Earth-object (NEO)
source population's orbital elements, and their geocentric orbit
residence-time distributions.  The NES population provides a test of
the NEO population statistics in a meteoroid size range that is not
well-sampled by contemporary asteroid surveys---there are only three
known NEOs with $H>32$ as of Oct 22, 2011---and could provide a remote
laboratory for detailed long-term studies of the physical properties
of the smallest asteroids. The long-term future concept of a
spacecraft mission to retrieve an entire meteoroid from Earth orbit
would provide an unprecedented scientific opportunity.

\begin{table}[!h]
\centering
\begin{tabular}[c]{ll}
\hline
NES & Natural Earth Satellite \\
SFD & Size-Frequency Distribution \\
NEO & Near-Earth Object \\
EMS & Earth-Moon System \\
TP  & Test Particle \\
TC  & Temporary Capture \\
TCO & Temporarily-Captured Orbiter \\
TCF & Temporarily-Captured Flyby \\
ISP & Intermediate Source Population \\
SP  & Source Population \\
\hline
\end{tabular}
\caption{List of acronyms.}\label{table:acronyms}
\end{table}

Despite a large body of work on satellite capture by the gas giants,
mainly Jupiter and Saturn, there has been surprisingly little
published about the Earth's natural satellites other than the
Moon. The origin and evolution of the population of
temporarily-captured irregular natural Earth satellites (NES) is
entirely unknown.

To the best of our knowledge, the earliest paper mentioning NESs other
than the Moon was \citeasnoun{1913JRASC...7..145C}. He explained that
the great meteor procession witnessed in North America on Feb 9, 1913
``had been traveling through space, probably in an orbit about the
sun, and that on coming near the earth they were promptly captured by
it and caused to move about it as a satellite.'' A few years later
\citeasnoun{den1916a} concluded that ``the large meteors'' that passed
over Northern America in 1913 must have been temporary Earth
satellites because they traveled 2600 miles in the atmosphere
suggesting that the orbits were ``concentric, or nearly concentric,
with the Earth's surface.''

Since about 1957 a large body of work has been carried out in the
field of dynamics of low-altitude satellites with the main application
being artificial spacecraft. \citeasnoun{bak1958a} considered the
possibility that artificial satellites ``may be accompanied in their
journey through space by certain "natural" satellites.'' He
hypothesized that the heliocentric orbits of Earth-grazing meteors
would become geocentric and elliptical as a result of atmospheric
drag. In the next decade, \citeasnoun{cas1965a} suggested that the line
of craters and hexahedrite meteorites associated with the Campo del
Cielo craters in Argentina originated in a high-altitude break-up of a
temporary Earth satellite in a decaying orbit. They also hypothesize
that the North Chilean hexahedrites may be fragments of the same body
which made one more revolution before coming to ground.

One line of studies on the dynamics of NESs was motivated by the
hypothesis that the Moon would be the origin of the terrestrial
tektites but a nontrivial transport mechanism was needed to explain
their uneven distribution on the Earth's surface
\citeaffixed{oke1961a}{see, e.g.,}. They are now thought to originate
in terrestrial impact events.

\citeasnoun{cli1979a} estimated the maximum speed of an object that
can be captured by the Earth-Moon system (EMS) if the object has a
very close encounter with the Moon. He concluded that in the four-body
system (Sun-Earth-Moon-asteroid) an asteroid on a heliocentric orbit
may be captured through a very close encounter with the Moon.

The first identification of an NES was only in the last decade.  While
\citeasnoun{bag1969a} provided direct and indirect evidence for
electrically charged NESs, \citeasnoun{mee1973a} later revealed
elementary misunderstandings and contradictions in the earlier work
that completely refute any evidence of NESs. \citeasnoun{tan1997a}
discussed the origin of 1991 VG and suggested that this recurrently
temporarily-captured object could be a piece of lunar ejecta formed by
a large impact. The estimated absolute magnitude has a wide range
($26.7 < H < 29.0$) even when just considering slope parameters
typical for natural objects with high albedos. However, the
possibility that 1991 VG could be artificial has not yet been ruled
out although this is an unlikely scenario due to the large projected
area (G. Tancredi, personal communication). Interestingly, 1991 VG is
currently flagged as a Virtual Impactor by the Jet Propulsion
Laboratory's impact-monitoring system, SENTRY. \citeasnoun{kwi2009a}
presented photometry of asteroid 2006 RH$_{120}$ ($H=29.9\pm0.3$)
which orbited the Earth within the Earth's Hill sphere from July 2006
until July 2007.  It has been concluded that it can not be a man-made
object based on its low area-to-mass ratio (P.\ Chodas, personal
communication) and high circular-polarization ratio and low albedo
from radar observations (L.\ Benner, personal communication). The
preliminary estimate for its diameter is $>2.3\meter$ based on
continuous-wave radar measurements.

We have not tested NEO orbits in the Minor Planet Center catalogue for
temporary capture by the EMS, but have assumed that if such objects
would exist they would have been reported in other publications. E.g.,
\citeasnoun{kwi2009a} give an extensive and detailed list of NES
candidates, and argue that 2006 RH$_{120}$ is the only object
certainly known to be a NES. Testing for past or future captures is
not trivial --- assuming that astrometric measurements are only
available after the escape in the previous case --- because the
orbital uncertainties and long extrapolation intervals combined with
the fractal nature of the orbit distribution for capturable objects
\cite{mur1989a} might lead either to missing some captures or to false
positive captures. The utility of adding possible NES candidates would
thus be questionable for the purpose of estimating the validity of our
results on the NES SFD. On the other hand, if there are observations
of an object both before and after a capture it is likely that there
also are some observations (or observers have at least attempted to
get them) during the capture. It is likely that these events or their
mere possibility would have been published in the literature.

The Earth's quasi-satellites have some common characteristics with the
NESs. (For a description of their dynamics see, e.g.,
\citeasnoun{mik2006a} and references therein.) The essential
difference between satellites and quasi-satellites is that the NES
orbit depends critically on the gravity of the EMS while the orbit of
a quasi-satellite would hardly change if the EMS suddenly ceased to
exist because it is orbiting the Sun on an Earth-like orbit in the
vicinity of the Earth. Quasi-satellites thus form a subgroup of
objects with co-orbital motion with respect to planets (objects on
tadpole and horseshoe orbits form the other two subgroups). Based on
the NEO model by \citeasnoun{bot2002a}, the steady-state population of
Earth's co-orbitals was estimated by \citeasnoun{mora2002a} to be
around 13--19 objects with $H<22$ and the length of the co-orbital
motion episodes range from 25~kyr to 1~Myr in their integrations. The
origin and evolution of Earth's known co-orbitals has recently been
studied by \citeasnoun{bra2008a} who conclude that these objects
should exist on Earth-like orbits for around 10kyr (note the
discrepancy with \citename{mora2002a}) --- three orders of magnitude
shorter than the average lifespan of an NEO.

The general approach we have chosen to determine the NES SFD is to
first calculate the capture probability as a function of the source
population's heliocentric orbital elements and then combine the
probabilities with the best available NEO population models
\cite{bot2000a,bot2002a,bro2002a}. The questions we will answer are:
\begin{itemize}

\item What is the steady-state population of temporary-captured NESs?

\item What are the pre- and post-capture orbit distribution for NESs?

\item How long do capture events last?

\item How many orbits does an NES typically complete before escaping
  the EMS?

\item What fraction of NESs impact the Earth and Moon while captured?

\item What are the orbit characteristics of an NES?

\item Can an NES become a temporary Moon satellite?

\item Are capture events equally likely to happen throughout the year?

\end{itemize}

\section{Definitions and methods}
\label{sec:theory}

\subsection{Definition for temporary capture}
\label{sec:definition}

Following \citeasnoun{kar1996a}, we use a two-fold definition for a
temporary satellite capture by a planet (or any object orbiting the
Sun including the Moon) by requiring simultaneously that
\begin{enumerate}
\item the planetocentric Keplerian energy $E_\mathrm{planet} < 0$, and
  that
\item the planetocentric distance is less than three Hill radii for
  the planet in question (e.g., for the Earth $3R_{H,\oplus} \sim
  0.03\AU$).
\end{enumerate}
We call a test particle (TP) a temporarily-captured orbiter (TCO) if
it makes at least one full revolution around the planet in a
co-rotating frame while being captured (the line from the planet to
the Sun is fixed in this coordinate system). If a temporarily-captured
TP fails to complete a full revolution around the planet we call it a
temporarily-captured fly-by (TCF). 1991 VG was neither a TCF nor a TCO
during its 1991-1992 encounter \citeaffixed{tan1997a}{due to energy
  and/or distance criteria; see Fig 1 in} whereas 2006 RH${}_{120}$
was a TCO during its 2006-2007 encounter. The practical limit to the
minimum duration of a TCF is two integration steps because a single
integration step does not allow us to estimate the capture duration.

We count the number of revolutions by recording the longitudinal angle
traversed during the capture. The longitude is measured in a
co-rotating ecliptic coordinate system where the line connecting the
planet and the Sun forms the line of reference. At every timestep we
add the difference between the current and the previous longitude to a
counter which results in a negative angle for retrograde orbits (as
seen from the north ecliptic pole) and a positive angle for prograde
orbits. A horseshoe-type orbit would result in less than one apparent
revolution being recorded regardless of how many loops the object
completes. 2000 SG${}_{344}$ did not qualify as a TCO during its Earth
encounter in 2000.

\subsection{Generation of initial conditions for test particles} 
\label{sec:initialconditions}

In what follows we describe the technique we use to generate initial
orbits for TPs in a volume that harbors NEOs shortly before their
capture by the EMS. The fundamental idea is that TPs that can get
captured by the EMS are constrained to a fairly small volume in the
heliocentric orbital elements (semimajor axis $a_\mathrm{h}$,
eccentricity $e_\mathrm{h}$, inclination $i_\mathrm{h}$) space
centered around the Earth's orbit.  We will call the population of
capturable NEOs the intermediate source population (ISP).

We start by drawing a random
($a_\mathrm{h}$,$e_\mathrm{h}$,$i_\mathrm{h}$) triplet from a uniform
distribution just slightly larger than the volume harboring
``capturable'' orbits ($0.87\AU < a_\mathrm{h} < 1.15\AU$, $0 <
e_\mathrm{h} < 0.12$, and $0\deg < i_\mathrm{h} < 2.5\deg$).  We then
draw a random longitude of node, $\Omega_\mathrm{h}$, argument of
perihelion, $\omega_\mathrm{h}$, and mean anomaly, $M_{0\mathrm{h}}$
triplet each from a uniform distribution in the interval $[0,2\pi|$
  radians. The intervals for
  ($a_\mathrm{h}$,$e_\mathrm{h}$,$i_\mathrm{h}$) were selected
  empirically to ensure that they span the complete phase space of
  initial conditions that can result in temporary captures. In order
  to average over different geometries between the asteroid, the
  Earth, the Moon, and the Sun, the epoch is randomly selected from a
  uniform distribution spanning the $\sim$19-year Metonic Cycle. The
  Metonic Cycle is the period over which a given lunar phase repeats
  at the same time of the year and it is accurate to a couple of
  hours.

We add the generated trial orbit to the sample of TPs to be integrated
thru the EMS if
\begin{enumerate}
\item the TP's geocentric distance is between 4 and 5 Earth's Hill
  radii (0.04--0.05$\AU$) at the epoch,
\item the TP has a slow-enough geocentric speed
  $v_\mathrm{g}(r)<v_e(r)+2.5\km\second^{-1}$ where $v_e(r)$ is the
  escape speed at geocentric distance $r$, and
\item the direction angle $\theta<130\deg$, the angle between the
  instantaneous geocentric velocity vector and the instantaneous
  TP-centric position vector of the Earth (see
  Fig.\ \ref{fig:geometry}).
\end{enumerate}
In Sect.\ \ref{sec:results_init} we will explicitly verify that the
cuts imposed on the distribution of integrated TPs do not reject
candidates that might evolve into TCOs on this pass thru the EMS. All
trial orbits are counted and the result stored so that we can
calculate the capture probability as a function of
($a_\mathrm{h}$,$e_\mathrm{h}$,$i_\mathrm{h}$).

This `brute-force' technique is feasible in our case only because the
likelihood of a close Earth encounter is highest for Earth-like orbits
with $a_\mathrm{h}\sim1\AU$, $e_\mathrm{h}\sim0$, and
$i_\mathrm{h}\sim0$. Approximately one in 1,000 of the generated
random orbits (also known as state vectors) fulfill all
requirements. To generate $10^7$ TP orbits we thus need to generate
``only'' about $10^{10}$ trial orbits. Appendix
\ref{appsec:initialconditions} describes a generic technique that can
be used for generating orbits in any situation when the
($a_\mathrm{h}$,$e_\mathrm{h}$,$i_\mathrm{h}$) distribution and the
heliocentric position vector
$(x_\mathrm{h},y_\mathrm{h},z_\mathrm{h})_{t_0}$ distribution are
known.

\subsection{Integrations} 
\label{sec:integrations}

The orbit of each TP was integrated using the Gram, Bulirsch and Stoer
algorithm \cite{sto2002a} starting with the initial conditions
described in Sect.\ \ref{sec:initialconditions}. Only gravitational
perturbations from the point-like masses of the Sun, the eight planets
and the Moon were taken into account. We stress the fact that
atmospheric drag was not included in the integrations. We did not
integrate the perturbers' orbits but instead we obtained the positions
from a special version of IMCCE's INPOP planetary ephemerides
\cite{fie2008a}. The integrations were performed using between 8 and
1,024 cores on the Jade supercluster (SGI Altix ICE 8200) located at
the Centre Informatique National de l'Enseignement Sup\'erieur (CINES)
in France.

Each TP was integrated separately and for at least 2~kdays to allow
almost co-moving objects to approach the Earth.  TPs that were still
captured after 2~kdays were integrated until they escaped the EMS. The
integration was stopped if the TP collided with the Earth or the Moon
(i.e., its geocentric or lunacentric distance were $\le
4.25\times10^{-5}\AU$ or $\le 1.16\times10^{-5}\AU$ corresponding
roughly to the Earth's and Moon's radius, respectively). The orbital
elements of the TP with respect to both the Earth and the Moon were
computed at each integration step --- the length of which varied from
a fraction of a day to tens of days depending on the automated
optimization --- and stored if $E_\oplus < 0$ or
$E_\mathrm{\rightmoon} < 0$ (see Sect. \ref{sec:definition}).

\subsection{NES pre-capture heliocentric orbit-density distribution}
\label{sec:theory_od}

Let the source population's (i.e. NEOs') debiased SFD and heliocentric
orbit-density distribution be represented by $N_\mathrm{NEO}(H)$ and
$R_\mathrm{NEO}(a_\mathrm{h},e_\mathrm{h},i_\mathrm{h})$,
respectively.

The heliocentric orbit-density distribution for the subset of objects
in the ISP that will eventually evolve into the target population is
\begin{equation}\label{eq:orbitdensity1}
  R_\mathrm{ISP}'(a_\mathrm{h},e_\mathrm{h},i_\mathrm{h}) = E_\mathrm{gen}(a_\mathrm{h},e_\mathrm{h},i_\mathrm{h}) \times
  E_\mathrm{capt}(a_\mathrm{h},e_\mathrm{h},i_\mathrm{h}) \times R_\mathrm{NEO}(a_\mathrm{h},e_\mathrm{h},i_\mathrm{h})\,
\end{equation}
where $E_\mathrm{gen}(a_\mathrm{h},e_\mathrm{h},i_\mathrm{h})$ is the
efficiency for generating initial conditions to be integrated,
$E_\mathrm{capt}(a_\mathrm{h},e_\mathrm{h},i_\mathrm{h})$ is the
capture efficiency among the integrated orbits, and the prime
indicates that $R_\mathrm{ISP}'$ is a subset of the heliocentric
orbit-density distribution for the ISP, $R_\mathrm{ISP}$.

The generation efficiency is defined as
\begin{equation}
  E_\mathrm{gen}(a_\mathrm{h},e_\mathrm{h},i_\mathrm{h}) =
  \frac{N_\mathrm{int}(a_\mathrm{h},e_\mathrm{h},i_\mathrm{h})}{N_\mathrm{aei}(a_\mathrm{h},e_\mathrm{h},i_\mathrm{h})}
\end{equation}
where $N_\mathrm{int}(a_\mathrm{h},e_\mathrm{h},i_\mathrm{h})$ is the
number of TPs that were integrated (see
Sect.\ \ref{sec:initialconditions}) and
$N_\mathrm{aei}(a_\mathrm{h},e_\mathrm{h},i_\mathrm{h})$ is the total
number of trial orbits.

The capture efficiency of the integrated TP orbits is defined as
\begin{equation}
  E_\mathrm{capt,int}(a_\mathrm{h},e_\mathrm{h},i_\mathrm{h}) =
  \frac{N_\mathrm{capt}(a_\mathrm{h},e_\mathrm{h},i_\mathrm{h})}{N_\mathrm{int}(a_\mathrm{h},e_\mathrm{h},i_\mathrm{h})}\,,
\end{equation}
where $N_\mathrm{capt}(a_\mathrm{h},e_\mathrm{h},i_\mathrm{h})$ is the
number of captured TPs.

The orbit-density distribution can then be reduced to
\begin{align}\label{eq:orbitdensity2}
  R_\mathrm{ISP}'(a_\mathrm{h},e_\mathrm{h},i_\mathrm{h}) & =
  \frac{N_\mathrm{capt}(a_\mathrm{h},e_\mathrm{h},i_\mathrm{h})}{N_\mathrm{aei}(a_\mathrm{h},e_\mathrm{h},i_\mathrm{h})}
  \times R_\mathrm{NEO}(a_\mathrm{h},e_\mathrm{h},i_\mathrm{h}) \\
  & =
  E_\mathrm{capt}(a_\mathrm{h},e_\mathrm{h},i_\mathrm{h}) \times
  R_\mathrm{NEO}(a_\mathrm{h},e_\mathrm{h},i_\mathrm{h})\,.
\end{align}
Note that $E_\mathrm{capt}(a_\mathrm{h},e_\mathrm{h},i_\mathrm{h})$ is
defined only when
$N_\mathrm{aei}(a_\mathrm{h},e_\mathrm{h},i_\mathrm{h}) \ne 0$, which
sets an upper limit on the resolution of the binning assuming a fixed
number of test particles.  In practice, we want
$N_\mathrm{aei}(a_\mathrm{h},e_\mathrm{h},i_\mathrm{h}) \gg 1$ to
minimize the statistical error on the results. It is also important to
ensure that $N_\mathrm{aei}(a_\mathrm{h},e_\mathrm{h},i_\mathrm{h})$
is essentially constant over the relevant volume or, alternatively,
correct for an uneven distribution in
($a_\mathrm{h}$,$e_\mathrm{h}$,$i_\mathrm{h}$) space. In our case the
technique used for generating TP orbits ensures that the distribution
in ($a_\mathrm{h}$,$e_\mathrm{h}$,$i_\mathrm{h}$) space is uniform.

\subsection{Steady-state size-frequency distribution}
\label{sec:theory_sfd}

A relatively well-known result in statistical physics is that in the
steady-state scenario the size, $N$, mean lifetime, $\bar L$, and flux
rate, $F$, into (or out of) the population are related by
\begin{equation}\label{eq:steadystate}
N = F \bar L\,. 
\end{equation}
Thus, the NES steady-state SFD, $N_\mathrm{NES}(H)$, can be determined
from $F_\mathrm{NES}(H)$ and $\bar L_\mathrm{NES}$. The latter can be
obtained from orbital integrations by calculating the average time
that an object fulfills the conditions of being an NES. We determined
$F_\mathrm{NES}$ using two different methods as described in the
following two subsections.

\subsubsection{NES Flux determination \protect \citeaffixed{mora2002a}{following}}
\label{sec:ftrp1}

In this method the flux into the NES population is obtained from
\begin{equation}\label{eq:fluxtr1}
F_\mathrm{NES1}(H) = r_\mathrm{ISP} \, N_\mathrm{ISP}'(H)\,,
\end{equation}
where $r_\mathrm{ISP}$ is the fractional decay rate from the ISP into
the NES population and $N_\mathrm{ISP}'(H)$ is the number of objects
in the ISP that will eventually evolve into the NES population:
\begin{equation}\label{eq:isp_sfd}
  N_\mathrm{ISP}'(H) = N_\mathrm{NEO}(H) \iiint
  R_\mathrm{ISP}'(a_\mathrm{h},e_\mathrm{h},i_\mathrm{h}) \, \ud
  a_\mathrm{h} \, \ud e_\mathrm{h} \, \ud i_\mathrm{h}\,.
\end{equation}
The triple integral is a scaling factor that is directly proportional
to the capture efficiency and can be calculated by integrating over
Eq.\ \ref{eq:orbitdensity2}. Note that the steady-state SFD for the
intermediate source population $N_\mathrm{ISP}'(H)$ is not equal to
the steady-state SFD for the target population $N_\mathrm{NES}(H)$
because of the different average lifetimes in the two regions.

To calculate $r_\mathrm{ISP}$ we imagine that we suddenly stopped
feeding the ISP. The $N_\mathrm{ISP}'$ objects in the ISP would start
decaying into the NES population at a rate
\begin{equation}\label{eq:decayrate}
\frac{\ud N_\mathrm{ISP}'}{\ud t} = -r_\mathrm{ISP}(t) \, N_\mathrm{ISP}'\,.
\end{equation}
For a sufficiently short time interval at some time $t$ we can assume
that the fractional decay rate is constant,
$r_\mathrm{ISP}(t)=r_\mathrm{ISP}$, and integrate the equation to
obtain
\begin{equation}\label{eq:decayrate_integrated}
N_\mathrm{ISP}'(t) = N_\mathrm{ISP}' \, \exp{[-r_\mathrm{ISP}t]}\,,
\end{equation}
where $N_\mathrm{ISP}'(t)$ is the number of objects left in the ISP at
time $t$ that will eventually evolve into the NES population. The
fractional decay rate can thus be obtained by fitting a straight line
to $\ln{[N_\mathrm{ISP}'(t)]}$ over some short time interval during
which $r_\mathrm{ISP}$ can be assumed constant.  With both of the
right-hand terms in Eq.\ \ref{eq:fluxtr1} determined we can then
calculate $F_{NES1}$.

\subsubsection{ NES Flux determination (Alternative)}
\label{sec:ftrp2}

The second method for estimating the flux into the NES population is
based on the assumption that the flux from the source population into
the ISP, $F_\mathrm{ISP}(H)$, is proportional to the flux from the ISP
into the NES population, $F_\mathrm{NES}(H)$. The size of the
steady-state population in the ISP, $N_\mathrm{ISP}(H)$, can be
estimated using the NEO model and knowledge of how the TP orbits were
generated. The average lifetime of the objects in the ISP, $\bar
L_\mathrm{ISP}$, is measured from the TP integrations. The flux into
the ISP is then
\begin{equation}\label{eq:fluxis}
F_\mathrm{ISP}(H) = \frac{N_\mathrm{ISP}(H)}{\bar L_\mathrm{ISP}}\,. 
\end{equation}
From the integrations we can estimate the fraction
$f_\mathrm{NES/ISP}$ of objects in the ISP that eventually reach the
NES population and then the flux into the NES population as a function
of the absolute magnitude is
\begin{equation}\label{eq:fluxtr2}
F_\mathrm{NES2}(H) = f_\mathrm{NES/ISP} \, F_\mathrm{ISP}(H)\,.
\end{equation}

\section{Results and discussion}
\label{sec:results}

Unless otherwise specified this section applies to TCOs only.

\subsection{The nominal model and the barycentric model}

To investigate the role of the Moon in capturing temporary satellites
we used two different dynamical models: 1) the \emph{nominal model} in
which the Earth and Moon are treated as separate perturbers and 2) the
\emph{barycentric model} where the Earth and the Moon are combined
into a single perturber located at the EMS barycenter with the
combined mass of the Earth and the Moon ($\sim$1.012 Earth masses).

Changing the location of the Earth's mass from the geocenter to the
barycenter has a negligible effect on the geometry because the
difference between the geocenter and the EMS barycenter is only about
$3\times10^{-5}\AU$ or roughly a thousandth of the geocentric distance
to the TP's initial locations.  The apparent location of the Earth as
seen from an object at a geocentric distance of $0.04\AU$ will not
change by more than $\sim 0.043\deg$.

\subsection{Test particle generation and integration}
\label{sec:results_init}

\begin{table}[!h]
\centering
\begin{tabular}[c]{lr}
\multicolumn{2}{c}{{\bf Bulk properties of the generated}} \\
\multicolumn{2}{c}{{\bf test particles and captured objects}} \\
\hline
$N_\mathrm{tot}$ & $9,346,396,100$\\
$N_\mathrm{int}$ & $10,000,000$ \\
\hline
\multicolumn{2}{c}{Nominal model} \\
\hline
$N_\mathrm{TCF,short}$ & $209,917$ \\
$N_\mathrm{TCF,long}$ & $23,771$ \\
$N_\mathrm{TCO}$ & $18,096$ \\
$\bar L_\mathrm{TC}$ & $(62.2\pm1.3)\days$ \\
$\bar \tau_\mathrm{TC}$ & $(0.383\pm0.059)\rev$ \\
$\bar L_\mathrm{TCO}$ & $(286\pm18)\days$ \\
$\bar \tau_\mathrm{TCO}$ & $(2.88\pm0.82)\rev$ \\
\multicolumn{2}{l}{Fraction of TCOs with} \\
$\qquad\tau_\mathrm{TCO} > 2.88\rev$ & 11\% \\
$\qquad\tau_\mathrm{TCO} > 5\rev$ & 3.4\% \\
$\qquad\tau_\mathrm{TCO} > 50\rev$ & 0.1\% \\
$\qquad\L_\mathrm{TCO} > 271\days$ & 26\% \\
$\qquad\L_\mathrm{TCO} > 365\days$ & 15\% \\
$\qquad\L_\mathrm{TCO} > 3650\days$ & 0.1\% \\
\hline
\multicolumn{2}{c}{Barycentric model} \\
\hline
$N_\mathrm{TCF,short}$ & $320,748$ \\
$N_\mathrm{TCF,long}$ & $34,843$ \\
$N_\mathrm{TCO}$ & $4,494$ \\
$\bar L_\mathrm{TC}$ & $(53.76\pm0.11)\days$ \\
$\bar \tau_\mathrm{TC}$ & $(0.21751\pm0.00037)\rev$ \\
$\bar L_\mathrm{TCO}$ & $(334.6\pm1.7)\days$ \\
$\bar \tau_\mathrm{TCO}$ & $(1.1280\pm0.0019)\rev$ \\
\hline
\end{tabular}
\caption{$N_\mathrm{tot}$ is the total number of generated TPs,
  $N_\mathrm{int}$ is the number of integrated TPs,
  $N_\mathrm{TCF,short}$ is the number of TCFs making less than half a
  revolution, $N_\mathrm{TCF,long}$ is the number of TCFs making more
  than half a revolution but less than one, $N_\mathrm{TCO}$ is the
  number of TPs making more than one revolution, $\bar L_\mathrm{TC}$
  is the average duration of a temporary capture, $\bar
  \tau_\mathrm{TC}$ is the average number of revolutions during a
  temporary capture, $\bar L_\mathrm{TCO}$ is the average lifetime of
  a TCO, and $\bar \tau_\mathrm{TCO}$ is the average number of
  revolutions made by a TCO during the time of
  capture.}\label{table:bulkresults}
\end{table}

A possible concern with this study might be that the generated TPs
that were integrated through the EMS do not span the entire range of
`capturable' orbits.  In this case our results would imply only a
lower limit to the TCO rate.  However,
Figs.\ \ref{fig:nesc_init_elements}-\ref{fig:nesc_init_direction}
illustrate that our generated TPs expand {\it beyond} the necessary
ranges.  Specifically, Fig.~\ref{fig:nesc_init_elements} shows that
the generated heliocentric ($a$,$e$,$i$) distributions for the TPs
that evolve into TCOs occupy a smaller volume than all the integrated
TPs. Figure \ref{fig:nesc_init_vel} shows that the initial geocentric
speed of the generated TPS spans a much wider range of values than
that of the objects that become TCOs which typically have $v_{g}
\lesssim 2.2\km\second^{-1}$.  Finally, the direction-angle
distribution at the generation epoch does not have a cut-off at
$\theta_\mathrm{init}=90\deg$ as might be naively expected, but
extends to $\theta_\mathrm{init} \sim 125\deg$ --- still below our
$\theta_\mathrm{init}<130\deg$ criterion in generating the TPs (see
Fig.~\ref{fig:nesc_init_direction}). The reason some TCOs that are
initially moving away from the Earth turn around and approach the
Earth is not that they are attracted by the Earth's gravity: even when
the gravity of all the planets and the Moon are excluded from the
integrations these initially outward-moving TCOs still appear to turn
around and move towards the Earth. Hence, the behavior is due to the
mutual geometry of the Earth's and the asteroids' orbits.  We see that
the $v_{g}$ distribution is shifted towards faster speeds in the
nominal model compared to the barycentric model while the distribution
of $\theta_\mathrm{init}$ is wider for the nominal model. In other
words, the Moon allows faster objects to be captured over a wider
range of initial direction angles.

About 20\% of all TCFs and TCOs are captured on multiple occasions
during the 2-kday integration and for 14 TPs the multiple events
include TCO-level captures. While these events represent an
insignificant fraction of the capture events, to avoid double counting
in the analysis we only take into account the first TCO capture for
any TP.  When a TP is captured and released to the NEO population it
again becomes part of the input NEO population.

Some bulk parameters for the generated particles and capture events
that lasted longer than one integration step are listed in Table
\ref{table:bulkresults}. Note that the distribution of the TCO
lifetimes and their number of revolutions have extremely long tails in
the nominal model (Fig.\ \ref{fig:tco_duration}). For example, the
longest capture event lasts $325,039\days$ (or about $1,200\times\bar
L_\mathrm{TCO}$) during which the the TP makes $14,801\rev$ around the
Earth (about $5,140\times\bar \tau_\mathrm{TCO}$).

Our TCO distribution is not biased by a truncation in the integration
time.  More than 99.9\% of all TCOs complete their first revolution
less than $\sim$400 days after the generation epoch
(Fig.\ \ref{fig:nesc_init_a_dt}) and all TPs were integrated for at
least 2~kdays. The TPs that were captured after 2~kdays were
integrated until they escaped.

\subsection{Capture probability and capture mechanism for temporarily-captured orbiters}
\label{subsec:capture_mechanism}

Figure \ref{fig:nesc_init_elements} shows a subset the initial orbital
elements ($a_\mathrm{h}$, $e_\mathrm{h}$, $i_\mathrm{h}$, and
longitude of perihelion $\varpi_\mathrm{h}$) of a representative
sample of all integrated TPs overplotted by those resulting in
TCOs. The most striking feature is the almost complete lack of TCOs
that are initially on orbits with $a_\mathrm{h}\sim1\AU$. This is
neither due to a too short integration time which would prevent the
slowest TPs to reach the target region
(Fig.\ \ref{fig:nesc_init_a_dt}) nor can it be explained by the
intrinsically higher impact probability for $a_\mathrm{h}\sim1\AU$
orbits because the impact probability is only of the order of
$10^{-4}$ and there should be plenty of objects replenishing the wake
left by the impactors
(Fig.\ \ref{fig:collisions_a_histo}). Furthermore, the gap in
semimajor axis cannot be explained by very-Earth-like orbits not
reaching the capture volume near $L_1$ or $L_2$, because the width of
the gap is independent of eccentricity. We think that the single most
important explanation for the gap in the heliocentric semimajor-axis
distribution is related to few-body dynamics: when a TP approaches the
EMS on a near-circular orbit with $a_\mathrm{h} \sim a_\oplus$ it
often follows either a horseshoe orbit or a quasi-satellite orbit and
cannot get close enough to the EMS for capture
\citeaffixed{mur1999a}{see, e.g., Chapter 3.13 in}. To verify this
hypothesis, we first examined the Jacobi constant as a function of
$a_\mathrm{h}$
\begin{equation}
C = \frac{1}{2}v^2 + U(x,y,z)\,,
\end{equation}
where both the speed $v$ and the potential $U(x,y,z)$ are given in the
rotating frame (Fig.\ \ref{fig:aC}).  We define two $C$-$a_h$ regions
in the figure: 'A' representing TPs that become TCOs and 'B'
representing TPs that do not become TCOs even though the objects in
'B' have the same value of the Jacobi constant as the TPs in
'A'.  The figure shows we can rule out the possibility that TPs with
$a_\mathrm{h} \sim a_\oplus$ have too small a $C$ to enter the EMS
through $L_1$ or $L_2$ because the range in $C$ is identical in both
the 'A' and 'B' regions and there are many TCOs in 'A'. We then
randomly selected 100 TPs from both regions with the additional
constraint that $i_\mathrm{h}<1\deg$, and integrated the subsets of
TPs for $500\yr$. In region 'A' we find evidence for one horseshoe
orbit whereas in region 'B' we find evidence for about 16
quasi-satellite orbits and 17 horseshoe orbits. Thus we conclude
that roughly one third of the TPs in region 'B' are prevented from
close encounters with the EMS.

Figure \ref{fig:ceff_tco_aei} shows the TCO capture probability,
$E_\mathrm{capt}(a_\mathrm{h},e_\mathrm{h},i_\mathrm{h})$, as a
function of their pre-capture heliocentric orbital elements. The wings
of the ($a_\mathrm{h}$,$e_\mathrm{h}$) distribution follow the
perihelion $q_\mathrm{h}=1\AU$ and aphelion $Q_\mathrm{h}=1\AU$ lines
indicating that a TP on an orbit allowing a grazing encounter with the
Earth's orbit may lead to the TP becoming a TCO whereas Earth-like
$a_\mathrm{h}\sim1\AU$ orbits in general are not necessarily
capturable even with fairly low eccentricities
\citeaffixed{1979aste.book..391C}{cf.}. The logarithmic scale in
Fig.\ \ref{fig:ceff_tco_aei} may be misleading since most of the
capture probability is focused in the 0.01~AU semi-major axis ranges
between 0.98-0.99~AU and 1.01-1.02~AU with a slight preference for a
non-zero eccentricity in the range 0.01-0.02. Although
Figs.\ \ref{fig:nesc_init_elements} and \ref{fig:ceff_tco_aei} suggest
that the highest capture probability is achieved when $q_\mathrm{h}
\sim q_\oplus=0.983\AU$ or $Q_\mathrm{h} \sim Q_\oplus=1.017\AU$ and
$\varpi_\mathrm{h} \sim \varpi_\oplus$ --- that is, when line of
apsides between the Earth and the TP are aligned and their perihelion
and aphelion distances are similar thus resulting in co-linear
velocity vectors for the Earth and the TP --- the reality is more
complicated. Figure \ref{fig:ceff_tco_dlperqQ} shows that the maximum
of the capture probability is bifurcated (or bimodal) in $q$ and $Q$
and extended in $\varpi_\mathrm{h}$. The reason for the bifurcation
is, to the best of our knowledge, currently unknown but we presume
that it is a feature of the elliptic four body problem.

When comparing the capture probability distribution with the orbital
elements of known NEOs in Fig.\ \ref{fig:ceff_tco_aei} bear in mind
that the capture probability is a function of the TP's orbital
elements when they have a geocentric distance of $0.04$--$0.05\AU$
whereas the orbital elements for the known NEOs in the figure have not
been filtered based on their geocentric distance. The orbital elements
of the known NEOs would change through planetary perturbations if they
were integrated until they had similar geocentric distances as the
TPs. Despite the slight inconsistency, the one confirmed former TCO,
2006 RH$_{120}$, still lies in a region that has a non-zero capture
likelihood so that it could be captured again in the future. 

A comparison with the barycentric model underscores the importance of
the Moon for the capture events as shown in
Fig.\ \ref{fig:ceff_tco_aei_bary}. The wings in the
($a_\mathrm{h}$,$e_\mathrm{h}$) distribution clearly require the
presence of the Moon but also the width in $a_\mathrm{h}$ of the
non-zero capture probability region increases dramatically when the
Moon is included in the integrations. Moreover, the results for the
barycentric model indicate that the presence of the Moon was of
critical importance in the capture of 2006 RH$_{120}$ because its
heliocentric elements prior to capture (on 2006 March 21 UT) at a
geocentric distance of about $0.047\AU$ ---
$a_\mathrm{h}\sim0.954\AU$, $e_\mathrm{h}\sim0.051$, and
$i_\mathrm{h}\sim0.565\deg$ --- would have placed it outside the
capture region for the barycentric model. The comparison between the
capture probability for the nominal and barycentric models indicates
that one should be wary of results that utilized a barycentric model
to analyze temporary satellite captures by the Earth
\citeaffixed{tan1997a}{c.f.,}.

The capture regions for slow-moving TCOs --- the region where
the geocentric orbital energy for TCOs first turns negative --- are in
the vicinity of the Sun-Earth Lagrange equilibrium points $L_1$ and
$L_2$ with roughly equal shares for both regions \citeaffixed[and
  references therein]{2010RAA....10..587B,2007MNRAS.377.1763I}{for
  illustrations and detailed explanations of the dynamics, see,
  e.g.,}. The $L_1$ and $L_2$ as derived for the circular restricted
three-body problem should only be understood as approximate reference
points when interpreting Fig.\ \ref{fig:rotcoord_at_capture} because
the fundamental problem with satellite captures by the EMS is the
elliptic four-body problem.

Retrograde geocentric orbits are preferred at the time of capture with
a share of approximately 2:1
(Fig.\ \ref{fig:rotcoord_at_capture}). Note that some 90\% of the
objects appear to move in a retrograde fashion in the rotating frame
due to their slow geocentric angular velocity and the fairly large
apparent angular velocity of the Sun. Not a single TP is
\emph{captured} by lowering the geocentric velocity via extremely
close lunar fly-bys so we conclude that even though the mechanism by
\citeasnoun{cli1979a} might be theoretically sound it is not important
in realistic capture scenarios. Among our integration results there
is, however, at least one example of a very close lunar fly-by
\emph{after} the initial capture that eventually lead to a very
long-lived TCO.

The volume of ($a_\mathrm{h}$,$e_\mathrm{h}$,$i_\mathrm{h}$)-space
harboring capturable TPs shrinks when the Moon is omitted from the
integrations in our barycentric model. Comparing
Figs.\ \ref{fig:nesc_init_elements} and
\ref{fig:nesc_init_elements_bary} shows that presence of the Moon
expands the semimajor-axis range while narrowing the inclination range
of the pre-capture TCO heliocentric orbits. Although conventional
wisdom suggests that the Moon's orbit is the primary reason for the lack
of long-lived TCOs, it is simultaneously the presence of the Moon that
increases the capture probability by allowing faster objects to be
captured \citeaffixed{cli1979a}{Table \ref{table:bulkresults};
  c.f.}. 

The longitude of perihelia for TCOs' pre-capture orbits are preferably
aligned with that of the Earth ($\sim103\deg$). We interpret the
Earth-like perihelion longitudes to be a geometric preference rather
than due to secular perturbations because the integrations typically
last for only a few years, too short a time interval for secular
perturbations to modify the heliocentric orbit distribution enough to
explain the results.  The longitude of perihelia preference manifests
itself as an inconstant TCO capture probability throughout the
year. Whereas the generation epochs are uniformly distributed over a
year the time-of-capture distribution has two annual peaks --- one in
late January and one in late July
(Fig.\ \ref{fig:tco_time_of_capture}). The peaks occur about 1--2
weeks after Earth's perihelion in January and aphelion in July (c.f.,
the alignment of the TCOs' longitude of perihelia with that of the
Earth's). The amplitude of the variation is $\lesssim 20$\% when the
data is binned with a resolution of about 7.3 days. Note that the
single verified TCO, 2006 RH$_{120}$, was, by our definition of a TCO,
captured in June. A typical TCO will thus get captured at Earth's
perihelion or aphelion and, since the distribution of the duration of
capture has a peak at 180~days (see Fig.\ \ref{fig:tco_duration}),
escape at the following Earth aphelion or perihelion, respectively. A
similar alignment of the capture and escape times has also been
observed for temporary satellite captures by Jupiter \cite{tan1990a}.

\subsection{Heliocentric orbit density for TCOs in the ISP}

To calculate the heliocentric orbit-density distribution
$R_\mathrm{ISP}'(a,e,i)$ for TPs in the ISP that will eventually
evolve into the NES population we use the \citeasnoun{bot2002a} NEO
model orbit distribution
$R_\mathrm{NEO}(a_\mathrm{h},e_\mathrm{h},i_\mathrm{h})$
(Fig.\ \ref{fig:tco_hodd}). A limitation with using this
$R_\mathrm{NEO}$ model is that its resolution is much lower than the
resolution we obtain for the raw capture probability $R_\mathrm{capt}$
--- we use 4 bins from the NEO model whereas the same volume contains
20,000 bins in $R_\mathrm{capt}$.  Thus, our knowledge of the
heliocentric pre-capture orbit-density distribution for TCOs is
limited by the lack of a high-resolution NEO orbit-distribution model.

\subsection{Steady-state size-frequency distribution of tem\-po\-ra\-ri\-ly-captured orbiters}
\label{subsec:steady-state}

\begin{table}[!hb]
\centering
\begin{tabular}[c]{lrl}
\hline
\multicolumn{3}{c}{{\bf Measured and published parameters}} \\
\hline
$C_\mathrm{NEO}$ & $13.26$ & \\
$n_\mathrm{ISP} \equiv N_\mathrm{int}$ & $10^7$ & \\
$|r_\mathrm{ISP}|$ & $(1.35575\pm0.00085)\times10^{-2}$ & $\days^{-1}$ \\
$R_\mathrm{ISP}$ & $4.0\times10^{-5}$ & \\
$\iiint R_\mathrm{ISP}'(a_\mathrm{h},e_\mathrm{h},i_\mathrm{h}) \, \ud
a_\mathrm{h} \, \ud e_\mathrm{h} \, \ud i_\mathrm{h}$ & $2.972\times10^{-10}$ & \\
$\bar L_\mathrm{ISP}$ & $(3.991\pm0.022)\times10^{1}$ & $\days$ \\
\hline
\multicolumn{3}{c}{{\bf Derived parameters}} \\
\hline
$N_0=N(H=24)$ & $(9.3874\pm4.7553)\times10^{4}$ & \\
$f_\mathrm{ISP/NEO}$ & $(1.06993\pm0.00034)\times10^{-3}$ & \\
$N_\mathrm{ISP}(H=24)$ & $(4.02\pm0.54)\times10^{-3}$ & \\
$F_\mathrm{ISP}(H=24)$ & $(1.007\pm0.136)\times10^{-4}$ & $\days^{-1}$ \\
$f_\mathrm{TCO/ISP}$ & $(1.810\pm0.013)\times10^{-3}$ & \\
$F_\mathrm{TCO1}(H=24)$ & $(3.78\pm1.92)\times10^{-7}$ & $\days^{-1}$ \\
$F_\mathrm{TCO2}(H=24)$ & $(1.82\pm0.25)\times10^{-7}$ & $\days^{-1}$ \\
\hline
\end{tabular}
\caption{Bulk parameters of the NEO and TCO populations.
  $C_\mathrm{NEO}$ is the constant in the power-law SFD and has been
  estimated assuming that there are 66 NEOs in the size range
  $13<H_V<15$ by \protect \citeasnoun{bot2000a}, $n_\mathrm{ISP}$ is
  the number of TPs in the ISP, $r_\mathrm{ISP}$ is the fractional
  decay rate from the ISP to the TCO population, $R_{ISP}$ is the
  fraction of all NEOs that are in the region from which TPs are
  generated,$\bar L_\mathrm{ISP}$ is the average lifetime of a TP in
  the ISP, $N_0$ is the extrapolated number of NEOs with $H=24$,
  $N_\mathrm{ISP}(H=24)$ is the number of objects with $H=24$ in the
  ISP in the steady-state scenario, $f_{ISP/NEO}$ is the fraction of
  NEOs generated that qualify for the ISP, $F_\mathrm{ISP}(H=24)$ is
  the flux of objects with $H=24$ into the ISP, $f_\mathrm{TCO/ISP}$
  is the ratio of objects in the ISP that enter the TCO population,
  and $F_{TCO1}$ and $F_{TCO2}$ are the flux of objects into the TCO
  region from the ISP calculated using two different
  methods.}\label{table:ssparams}
\end{table}

The TCO steady-state SFD, i.e., the number of TCOs as a function of
their absolute magnitude ($N_{TCO}(H)$), is determined by the NEO SFD,
the flux into the TCO population, $F_\mathrm{TCO1} = F_\mathrm{NES1}$
and $F_\mathrm{TCO2} = F_\mathrm{NES2}$ as discussed in
sect.\ \ref{sec:theory_sfd}, and the average TCO lifetime $\bar
L_\mathrm{TCO} = \bar L_\mathrm{NES}$ (obtained directly from the
integrations). Thus, we require an independent measurement of the NEO
SFD in order to determine the TCO SFD.

Motivated by the fact that the only verified TCO, 2006 RH$_{120}$, was
a few meters in diameter we expected that it was important to use a
NEO SFD relevant to meter-scale objects ($H\sim30$).  We used two of
the available SFDs, \possessivecite{rab2000a} $N_\mathrm{Ra00}(H)
\propto 10^{0.7H}$ valid for $24 < H < 31$, and
\possessivecite{bro2002a} estimate for Earth impactors
$N_\mathrm{Br02}(H) \propto 10^{(0.540\pm0.016)H}$ valid for $22 < H <
30$. Another independent estimate for the NEO SFD comes from the
analysis of the lunar impactors \citeaffixed{wer2002a}{see, e.g.,}
which is about an order of magnitude below the estimate for NEOs with
$H\sim30$ by \citeasnoun{rab2000a} but agrees well with the estimate
by \citeasnoun{bro2002a}.

Considering that we used the \citeasnoun{bot2000a} model for the NEO
orbit distribution we also used their SFD of $N_\mathrm{Bo00}(H)
\propto 10^{(0.35\pm0.02)(H-H_0)}$ which is strictly valid for $15 < H
< 22$ but we simply extrapolated it to the size range of interest.  At
$H=33$ the raw models differ in the predicted number of objects by
several orders of magnitude.  In order to bridge the gaps in $H$
between the different SFDs, we scale the absolute number of objects in
the source population (that is, in the NEO population) to
$N_0=N(H=24)$ which is obtained through extrapolation of the
\citeasnoun{bot2000a} model $N_\mathrm{Bo00}(H=24) =
C_{NEO}10^{(0.35\pm0.02)(H-13.0)} =
13.26\times10^{(0.35\pm0.02)\times11.0} = 93874\pm47553$
\citeaffixed{jed2003a}{cf.}. It turns out that the differences between
using the absolute numbers by \citeasnoun{rab2000a} and the broken
power law described above are negligible compared to the other error
sources; the flux at $H=30$ is different by a factor of about two. We
chose to use the broken power law because the SFD by
\citeasnoun{rab2000a} is not readily available in a functional form
but only as a plot.

We note that the constant $C_{NEO}$ was estimated 11 years ago when 53
NEOs in the size range $13<H<15$ were known \cite{bot2000a}. The
completeness level for that size range was assumed to be 80\% based on
the completeness level for NEOs with $H < 16$ which in turn was based
on data from 1996 to 1998. The total population of NEOs in the size
range $13<H<15$ was therefore assumed to be 66. During the past 11
years the number of NEOs with $H<15$ has grown to 55, but eight of
them have been discovered since April 2000 (the last one in
2004). These facts lead us to conclude that although some objects have
been discarded from the sample due to, for example, improved, fainter
absolute magnitudes, the total number of NEOs in the size range is
smaller than anticipated in 2000. The constant $C_{NEO}$ may thus be
up to about 20\% smaller than estimated by \citeasnoun{bot2000a}. We
will nevertheless use the original value for the constant since its
error is relatively small.

Table \ref{table:ssparams} lists all parameters that are used when
computing the fluxes and, further, the SFDs. The uncertainty estimates
for the parameters correspond to Gaussian 1-$\sigma$ limits which
have, in the case of fractions, been computed assuming that a binomial
distribution can be approximated by a Gaussian distribution in the
limit of large N. Correlations between the variables are not readily
available so for error propagation we have assumed that they are
uncorrelated.

At this point we stress that there are some important assumptions in
our calculation of the TCO SFD:
\begin{itemize}

\item The orbit-density distribution is assumed to be independent of
  the SFD.  While this assumption is probably valid for large NEOs it
  is certainly not true for the meter-class and smaller NEOs for which
  non-gravitational forces play an important role in the dynamical
  evolution of their orbits.  However, since no debiased NEO orbit
  distribution exists for the small NEOs we have no option but to make
  this assumption.  On the other hand, we expect that this work will
  provide constraints on the dynamical evolution of small NEOs and
  their orbit distribution once techniques are developed for
  identifying large numbers of TCOs.

\item The fraction of NEOs on Earth-like orbits is assumed to be
  exactly known because uncertainty estimates are not provided with
  the orbit-density distribution by \citeasnoun{bot2002a}. When
  estimating the integral in Eq.\ \ref{eq:isp_sfd} we also assumed
  that the capture efficiency as a function of orbital elements is
  known exactly.

\item The measured SFDs for the NEO source population extend to only
  $H\sim34$ so our calculation of the TCO SFD to smaller sizes is pure
  extrapolation.

\end{itemize}

To calculate the TCO SFD we first need to determine the fractional
decay rate, $r_\mathrm{ISP}$, for the first flux calculation,
$F_\mathrm{TCO1}$, (see Sect.\ \ref{sec:ftrp1}). To this end, we
calculate the time $t$ it takes for each TCO to get from a geocentric
distance of 0.05\,AU to the point of capture (i.e. correcting for the
fact that the TPs were generated in a shell at geocentric distances
from 0.04 to 0.05\,AU) and plot the natural logarithm of the number of
TPs still remaining to be captured as a function of $t$
(Fig.\ \ref{fig:decay}). The fractional decay rate is the slope of
this distribution --- but the slope is not constant so the measured
fractional decay rate will depend on the time interval over which the
slope is measured.  Instead, we measured the running slope over each 7
day interval and then calculating the number-weighted slope as the
weighted average with weights equal to the number of objects in each 7
day interval yielding
\begin{equation*}
F_\mathrm{TCO1}(H=24)=(3.78\pm1.92)\times10^{-7}\days^{-1}\,.
\end{equation*}

For the second method of calculating the TCO flux we need to know the
number of objects in the ISP, $N_\mathrm{ISP}$, with $H=24$, the
average lifetime in the ISP, $\bar L_\mathrm{ISP}$, and the fraction
$f_\mathrm{TCO/ISP}$ of TPs in the ISP that eventually become TCOs
(see Sect.\ \ref{sec:ftrp2}). $N_\mathrm{ISP}(H=24) = N_0 \times
R_\mathrm{ISP} \times f_\mathrm{ISP/NEO} \sim
(4.02\pm0.54)\times10^{-3}$ while the average lifetime in the ISP,
$\bar L_\mathrm{ISP}$, and fraction of TPs in the ISP that eventually
become TCOs, $f_\mathrm{TCO/ISP}$, are obtained directly from the
integration results; $(3.991\pm0.022)\times10^{1}\days$ and
$(1.810\pm0.013)\times10^{-3}$, respectively. The flux of TCOs using
the second method is then
\begin{equation*}
F_\mathrm{TCO2}(H=24)=(1.82\pm0.25)\times10^{-7}\days^{-1}\,.
\end{equation*}
The agreement between $F_\mathrm{TCO1}$ and $F_\mathrm{TCO2}$ is
surprisingly good considering that the fractional decay rate needed
for $F_\mathrm{TCO1}$ is computed assuming a constant slope --- an
assumption that does not hold in this particular case because the
slope is variable for any reasonable time interval.

Figure \ref{fig:tco_steady-state} shows that the TCO steady-state SFD
depends strongly on the assumed NEO source population's SFD and only
weakly on our method of calculating the TCO flux. There is
approximately 3 orders of magnitude difference in the predicted number
of TCOs larger than one meter in diameter ($H\lesssim 32$).

We can use the one confirmed TCO (2006 RH$_{120}$) with a diameter of
a few meters to discriminate between the three NEO population models
keeping in mind that no knowledge of 2006 RH$_{120}$ was used in our
TCO modeling.  In the steady state the \citeasnoun{rab2000a},
\citeasnoun{bro2002a}, and \citeasnoun{bot2000a} models predict that
the largest TCO always present in the steady-state population is
$\sim$3-m, $\sim$1-m and $\sim$0.2-m in diameter, respectively.  If
there really were one 3-meter diameter in orbit at any time we would
expect for many more of these objects to be known.  Similarly, the
\citeasnoun{bot2000a} model suggests that the time interval between
TCOs in 2006 RH$_{120}$'s size range is $> 100$~years implying that
the odds of finding such an object are small. Thus, the NEO SFD model
that is most consistent with the observed TCO distribution is the
\citeasnoun{bro2002a} model.

\subsection{Orbit characteristics and residence-time distributions
  for temporarily-captured orbiters}

Figures \ref{fig:typical_tco_xyz} and \ref{fig:typical_tco_kep}
illustrate that the TCO's osculating geocentric orbital elements
change dramatically on short timescales. A geocentric two-body orbit
is not adequate for describing the motion of TCOs even for relatively
short time periods.

The residence-time distributions shown in
Fig.\ \ref{fig:tco_residence_aei} were created by logging the time
that each TCO spends in each bin in the binned phase space and then
normalizing the distribution.  They can be thought of as the
instantaneous probability distributions for the orbital elements of
TCO population.  The distribution shows that NEOs can be captured and
evolve into almost any geocentric orbit although
low-$a_\mathrm{g}$-high-$e_\mathrm{g}$ orbits are strongly preferred
(note that the density scale is logarithmic). This can partly be
understood as a direct consequence of the fact that, in the geocentric
frame, all TCOs start and end with low-$a_\mathrm{g}$ and
high-$e_\mathrm{g}$ because they are on hyperbolic geocentric orbits
prior to capture and after escape. However, since
Fig.\ \ref{fig:tco_residence_gt5rev_aei} shows that low-$a_\mathrm{g}$
and high-$e_\mathrm{g}$ orbits are prominent in the residence-time
distributions for TCOs completing more than five revolutions,
low-$a_\mathrm{g}$ and high-$e_\mathrm{g}$ orbits are preferred
throughout long capture events. Note also that retrograde orbits are
slightly preference over prograde orbits.

The inclination distribution has a spike at
($a_\mathrm{g}\sim0.001\AU$,$i_\mathrm{g}\sim35\deg$) that is due to
only a handful of objects that have lifetimes up to 1000x longer than
the average TCO. These very long-lived TCOs evolve through, for
example, multiple lunar close approaches, into orbits with apogees
within the Moon's orbit. They are affected by the Kozai resonance
\cite{koz1962a} as revealed by Kozai synchronous oscillations in $e$
and $i$ (Fig.\ \ref{fig:xyGf_qQplot_mjdlt62000}) and the libration of
the argument of perigee around $\omega=270\deg$
(Fig.\ \ref{fig:xyGf_polarplot}). These orbital features prevent
long-lived TCOs from having close encounters with the Moon on short
timescales.

We did not discover any temporary lunar satellites fulfilling the TCO
criteria. We did find cases were the lunacentric Kepler energy was
negative for a short period of time but those TPs were not captured
long enough to complete a full revolution around the Moon.

\subsection{Terrestrial and lunar impacts by temporarily-cap\-tured orbiters}

About 1\% --- 169 out of 18,096 --- of all TCOs in our sample impacted
the Earth while being captured. The impact probability increases with
the duration of the capture so that about 18\% of TCOs with lifetimes
longer than 6 years eventually impact the Earth. The opportunity of
observing meteors and finding meteorites subsequent to the TCO phase
suggests that spectrometric observations of these bodies will maximize
their scientific return \citeaffixed{jen2009a}{cf.}.  In an additional
20 other cases (0.1\% of all TCOs) a terrestrial impact happened more
than one week after the TCO had escaped the EMS but within about one
to three years from the generation epoch.

A comparison of the rate of TCO-impactors to the background population
of impactors is nontrivial because only slow-moving objects were
integrated in the present work and TCOs occupy a smaller volume of the
phase space than non-TCOs (Fig.\ \ref{fig:collisions_a_dt}). However,
a rough comparison can be done if we limit consideration to the
slow-moving population and assume that the phase-space-volume
differences can be neglected by concentrating on two volumes harboring
TCOs. A limited volume of the phase-space is defined as $e_\mathrm{h}
< 0.035$, $i_\mathrm{h} < 1\deg$ and $0.97\AU< a_\mathrm{h} <0.995\AU$
or $1.005\AU< a_\mathrm{h} <1.03\AU$. This limited volume contains
575,100 TPs to be integrated and out of those 14,909 become
TCOs. Terrestrial impacts occur for 1,114 TPs and 104 of those were
TCOs. The TCO impact rate in the constrained phase space is thus
$104/14909\approx0.7$\% whereas the fraction of non-TCOs on similar
generated orbits producing impacts is $(1114-104)/(575100-14909)
\approx 0.2$\%. In other words, TCOs are nominally 3.5x more likely to
impact the Earth than non-TCOs on initially similar heliocentric
orbits.

The fraction of all Earth-impacting objects that were TCOs is
surprisingly high. In our integrations 189 TCOs impact the Earth
during a time span of 19 years so the annual TCO Earth-impact rate
from TCOs is about $10\,\yr^{-1}$. (Note that the following reasoning
does not depend on the time span nor the number of TPs.) To determine
the diameter of objects corresponding to this impact rate we note that
the average annual unnormalized (that is, raw) number of TCOs is about
$952\,\yr^{-1}$ which means that the corresponding unnormalized
steady-state population is about 707 TCOs (after multiplying by the
average TCO lifetime).  Figure~\ref{fig:tco_steady-state} shows that
the normalized steady-state population corresponding to this number of
objects have a maximum absolute magnitude of $H\sim37$ or a minimum
diameter of about $10\cm$.  Thus, the sizes of the above-mentioned
annual TCO impactors are typically $10\cm$ because the size
distribution for the impactors must be heavily skewed towards the
smallest sizes. Since \citeasnoun{bro2002a} states that there are
about $10^4$ objects larger than $10\cm$ impacting the Earth every
year, about 0.1\% of all Earth impactors are TCOs.

The terrestrial impact-speed distribution is extremely narrow with a
peak at approximately the Earth's escape speed, $11.2 \km\second^{-1}$
(Fig.\ \ref{fig:collisions_vel_histo}).  Indeed,
Fig.\ \ref{fig:tco_residence_rv} shows that the TCO speed
distribution as a function of geocentric distance is essentially that
of objects falling towards the Earth after starting with almost zero
speed. The impact-angle\footnote{The acute angle between the
  trajectory and the perpendicular to the Earth's `surface' (nearly
  equivalent to the atmospheric impact 'surface') at the point of
  impact.} distribution shows that TCO impacts span essentially the
whole range of possible angles
(Fig.\ \ref{fig:collisions_impact_angle_histo}). Recall, however, that
the present work does not take into account non-gravitational effects
such as atmospheric drag which could change the distribution of impact
angles.  These predictions may provide a means to differentiate
between TCOs and other meteors in radar data although it is unclear to
us whether the signal from NESs is strong enough to overcome the
background flux from, e.g., other meteor populations or spacecraft
debris re-entering the atmosphere. The detection of slow-moving
meteors is also massively biased against. There are, however, recorded
meteors with apparent speeds less than Earth's escape speed but it is
not clear whether these objects are natural or whether the low speeds
are due to deceleration before detection (personal communication with
P.\ Brown). We will discuss various techniques for discovering TCOs in
 a forthcoming paper on their observational characteristics.

None of the TCOs impacted the moon so this class of events is
extremely rare. In fact, only one lunar impact occured during the
integrations of $10^7$ TPs suggesting that low-speed lunar impacts in
general are extremely rare.

\subsection{Heliocentric orbit characteristics after a temporary capture}

The TCOs escape the EMS using the same route as during the capture ---
through the L1 and L2 points. This is a direct consequence of the time
reversibility of the gravitational-capture event. We integrated each
TCO (that did not impact the Earth) $1\yr$ forward in time from the
last date that it had $e_\mathrm{g}<1$ to find out how its
post-capture orbit compared to its pre-capture orbit. We checked that
most TPs are sufficiently far from the Earth at this stage so that
perturbations by the Earth on their orbits can be neglected. The major
difference between heliocentric pre-capture and post-capture orbits
for TCOs is the widening of the gap at $a_\mathrm{h}\sim1\AU$
(Fig.\ \ref{fig:TCO_postcapture_elements}). Although this would seem
to indicate that it is essentially only the semimajor axes that get
pushed away from the $a_\mathrm{h}\sim1\AU$ line, the reality is a bit
more complex. Figure~\ref{fig:TCO_postcapture_delements} shows that
some TPs that were TCOs entered the EMS on orbits with
$a_\mathrm{h}>1\AU$ and leave on orbits with $a_\mathrm{h}<1\AU$, and
vice versa. There is also a clear zone of avoidance so that orbits
with $a_\mathrm{h}\sim1\AU$ are dynamically impossible to reach. The
fractional changes in eccentricity and inclination can be large and
behave much like the semimajor axes but without the `zone of
avoidance' --- that is, large eccentricities and/or inclinations tend
to get smaller and small eccentricities and/or inclinations tend to
get larger.

The fact that the volume and shape of the post-capture orbit
distributions matches the volume and shape of the pre-capture
distributions can, again, be understood as a direct consequence of the
time reversibility of gravitational-capture events. The similarity of
the distributions suggests that a single object may be captured on
several different occasions before planetary perturbations force it to
leave the volume harboring capturable orbits.

\section{Conclusions}
\label{sec:conclusions}

We provide the first estimate of the orbit and size-distribution for
temporarily captured natural irregular satellites of the Earth. We
predict that there is a one-meter-diameter or larger NEO temporarily
orbiting the Earth at any given time. The NEO orbit and SFD model is
currently the main factor limiting the accuracy of our predictions.
Given the orbit distribution of \citeasnoun{bot2002a} the
\citeasnoun{bro2002a} NEO SFD is consistent with the only known TCO of
natural origin, 2006 RH$_{120}$, while the \citeasnoun{bot2002a} and
\citeasnoun{rab2000a} SFDs are each off by about an order of
magnitude.

Our integrated TCO population will allow us to examine different
scenarios for their detection and to estimate, e.g., the average time
from discovery to an accurately known orbit. It seems plausible that
the long-lived TCOs could have stable enough orbits to allow
successful searches to be carried out in specific regions of the sky.
Once TCOs can be reliably and frequently identified early enough in a
capture event they create an opportunity for a low-cost low-delta-v
meteoroid return mission \citeaffixed{elv2011a}{cf.}. The scientific
potential of being able to first remotely characterize a meteoroid and
then visit and bring it back to Earth would be unprecedented.

%
%
%
%

\newpage
\section*{Acknowledgments}

We acknowledge the thorough reviews by A.\ Christou and an anonymous
reviewer as well as helpful discussions with B.\ Gladman and
P.\ Wiegert concerning TCO dynamics, P.\ Chodas and L.\ Benner
concerning 2006 RH$_{120}$, and P.\ Brown concerning detection of NES
meteors. MG's and RJ's work was funded by a grant from the NASA NEOO
program (NNX07AL28G). MG was also funded by grants \#130989, \#136132
and \#137853 from the Academy of Finland. JV thanks the CINES team for
the use of the Jade SGI supercluster and their help in parallelizing
the orbital integration code.

%
%
%
%

\newpage
\bibliographystyle{agsm}
\bibliography{asteroid,educational}

%
%
%
%

\newpage
\setlength{\unitlength}{2mm} 
\begin{figure}[H]
  \centering
  \begin{picture}(60, 40)
    \thicklines
    \put(10,10){\circle{4}} 
    \put(8,13)
        {Sun}
    \put(40.2,25.1){\circle{3}} 
    \put(33,25)
        {Earth}
    \put(40.2,28.6){\circle*{1}} 
    \put(37,30)
        {Moon}
    \put(45,18.75){\circle{2}} 
    \put(45,16)
        {TP}

    \put(10,10){\vector(2,1){30.1}} 
    \put(25,19)
        {$\vect{r}_\mathrm{e}$}
    \put(10,10){\vector(4,1){35}} 
    \put(26,12.5)
        {$\vect{r}_\mathrm{p}$}
    \put(45,18.75){\vector(-3,4){4.7}} 
    \put(38,20.5)
        {$\vect{r}_\mathrm{p-e}$}
    \put(45,18.75){\vector(1,2){4}} 
    \put(49,23)
        {$\dot{\vect{r}}_\mathrm{p-e}$}

    \qbezier(43.5,21)(45,22)(46,21)    
    \put(45,22)
        {$\theta$}

  \end{picture}
  \caption{Vector and angle definitions in the Earth-Moon-Sun-object
    system (not to scale).  The test particle's heliocentric position
    vector and geocentric velocity vector are represented by
    $\vect{r}_\mathrm{p}$ and $\dot{\vect{r}}_\mathrm{p}$,
    respectively. The Earth's heliocentric position vector is marked
    with $\vect{r}_\mathrm{e}$.  The direction angle $\theta$ is the
    angle between the test particle's geocentric velocity and
    geocentric direction vectors.}\label{fig:geometry}
\end{figure}
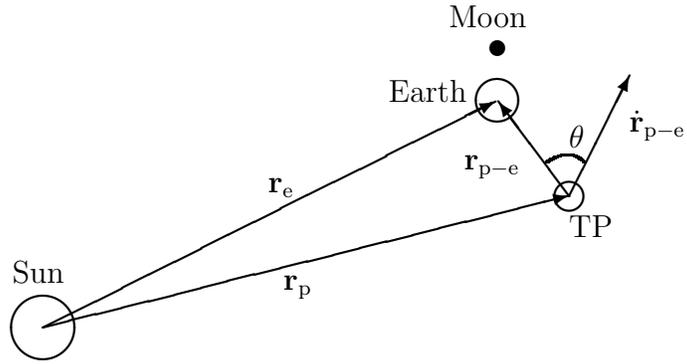

\newpage
\begin{figure}[H]
  \centering
  \includegraphics[width=0.75\textwidth]{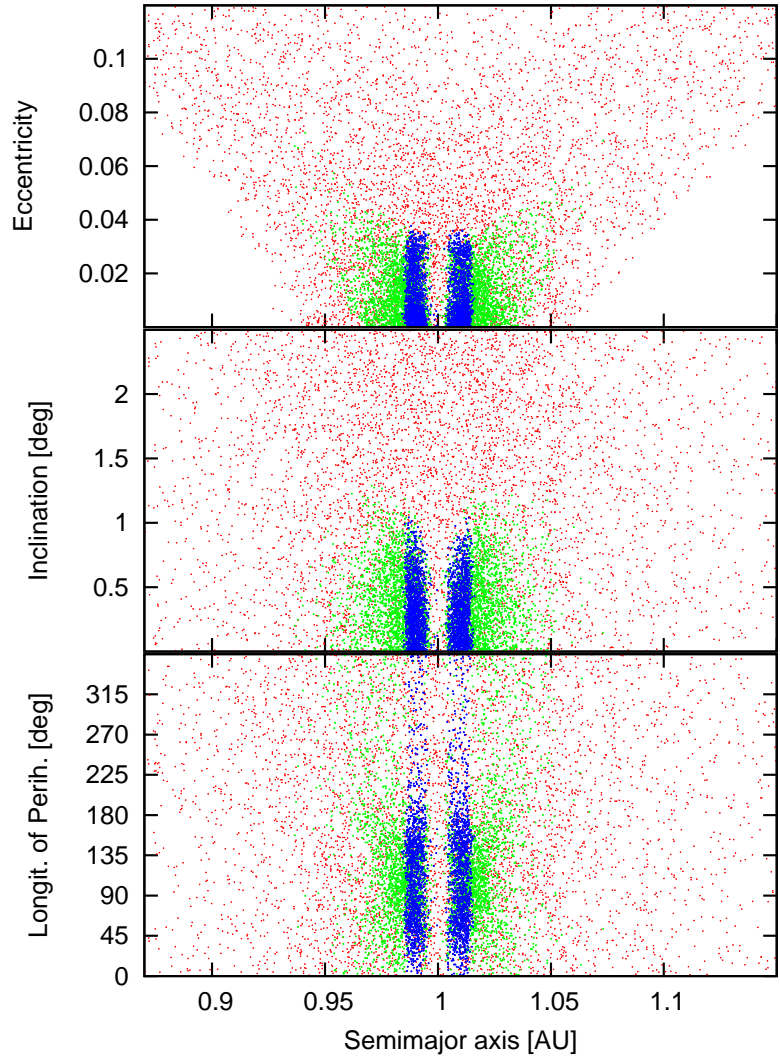}
  \caption{Representative samples of generated heliocentric Keplerian
    orbital elements for (red) all integrated test particles and
    temporarily-captured orbiters where the color indicates different
    intervals for the direction angle at the generation epoch: (green)
    $0\deg < \theta_\mathrm{init} < 90\deg$ and (blue) $90\deg <
    \theta_\mathrm{init} < 120\deg$.}\label{fig:nesc_init_elements}
\end{figure}

\newpage
\begin{figure}[H]
  \centering
  \includegraphics[height=\textwidth,angle=-90]{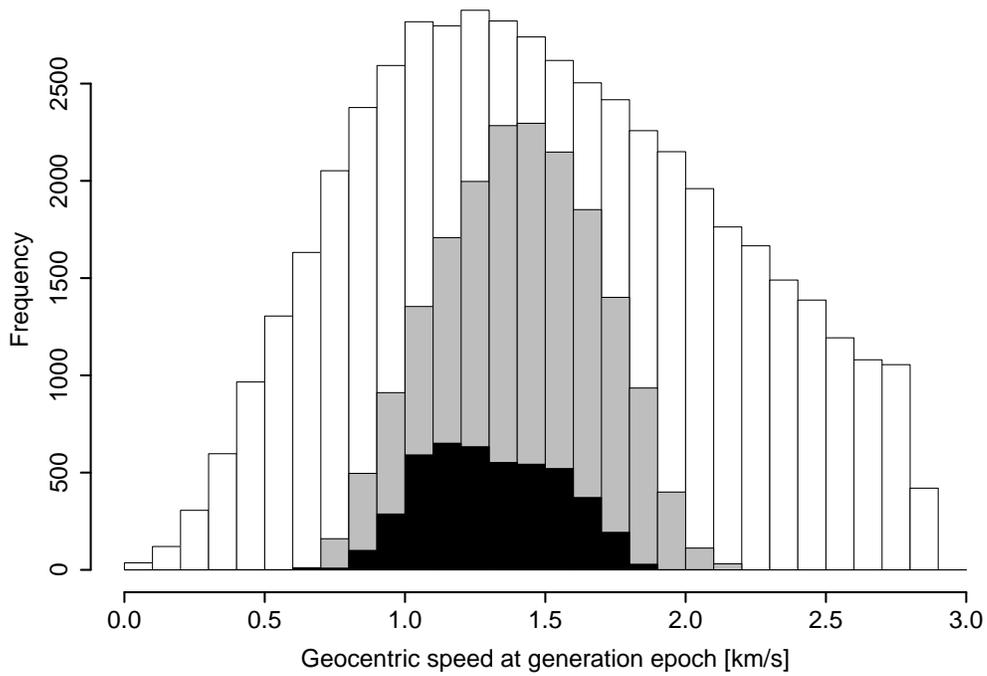}
  \caption{The geocentric speed distribution at the generation epoch
    for (white) 1/200th of all integrated test particles, (grey)
    temporarily-captured orbiters in the nominal model and (black) the
    barycentric model. Note that the mode of the distribution is
    smaller for the barycentric model.  }\label{fig:nesc_init_vel}
\end{figure}

\newpage
\begin{figure}[H]
  \centering
  \includegraphics[height=\textwidth,angle=-90]{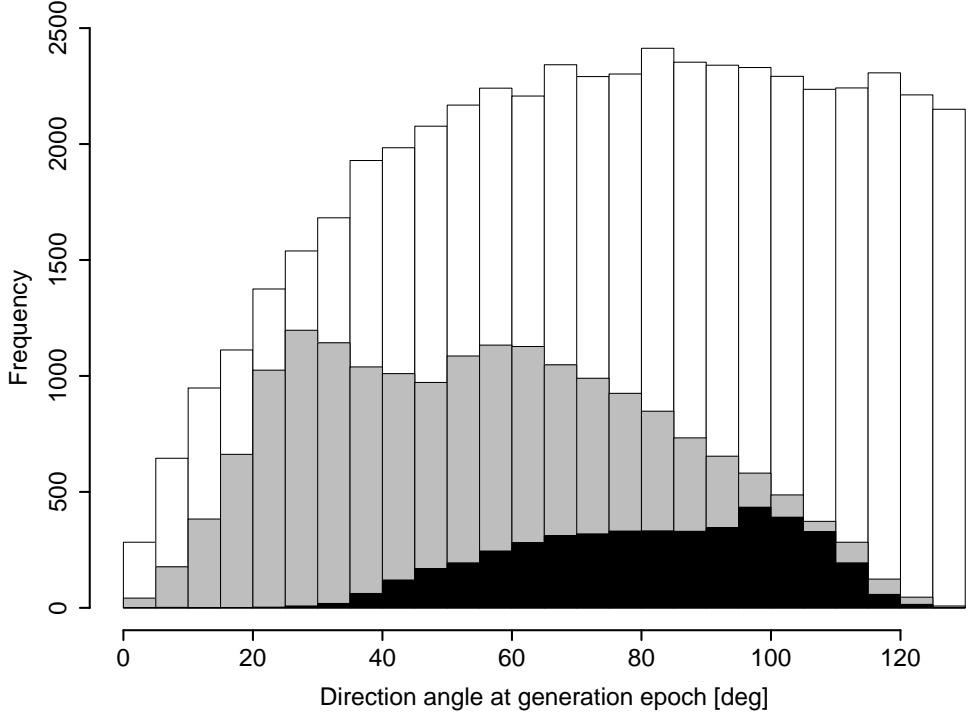}
  \caption{The distribution of direction angles at the generation
    epoch ($\theta_\mathrm{init}$) for the three groups of test
    particles in Fig.\ \ref{fig:nesc_init_vel}. At the initial epoch,
    at geocentric distances between 4 and 5 Hill radii, the TCO
    direction angles span a wide range of directions. Note that there
    are TPs that initially move away from the Earth
    ($\theta_\mathrm{init}>90\deg$) but are later captured. These
    objects typically have semimajor axis $\sim 1$
    (Fig.\ \ref{fig:nesc_init_elements}) and are not energetically
    bound to the EMS when they `turn around'. TCOs in the barycentric
    model are more likely to move perpendicular to the Earth than to
    approach it directly.}\label{fig:nesc_init_direction}
\end{figure}

\newpage
\begin{figure}[H]
  \centering
  \includegraphics[height=\textwidth,angle=-90]{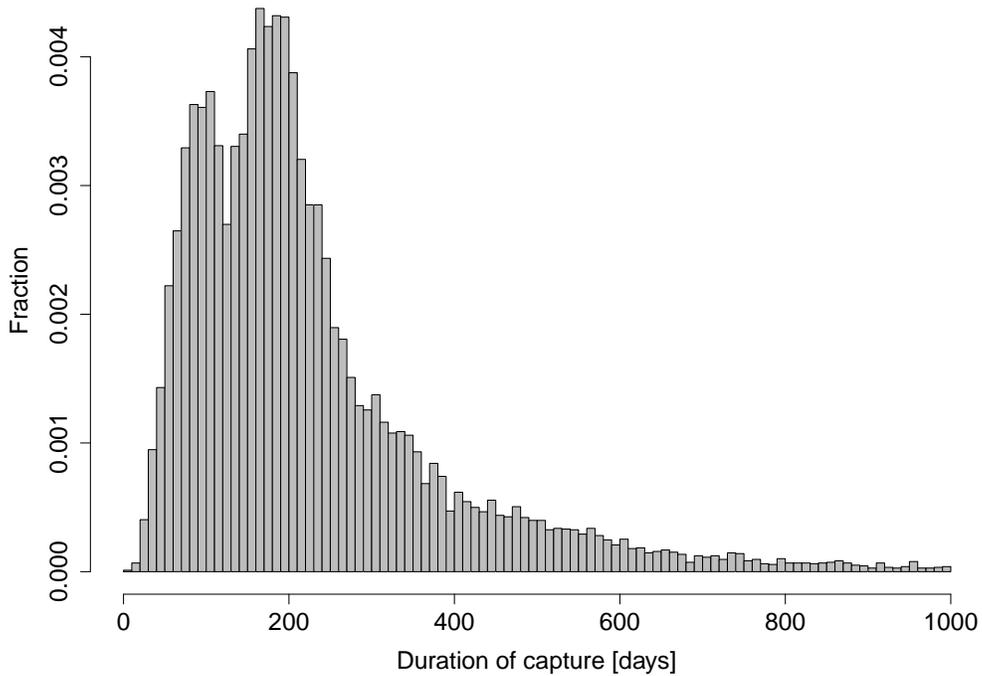}
  \caption{Duration of capture for temporarily-captured orbiters. The
    longest capture during the integrations lasted about 325,000 days
    but the histogram has been cut off at 1,000 days. The peaks are
    located at about 90 days and about 180 days.}\label{fig:tco_duration}
\end{figure}

\newpage
\begin{figure}[H]
  \centering
  \includegraphics[height=\textwidth,angle=-90]{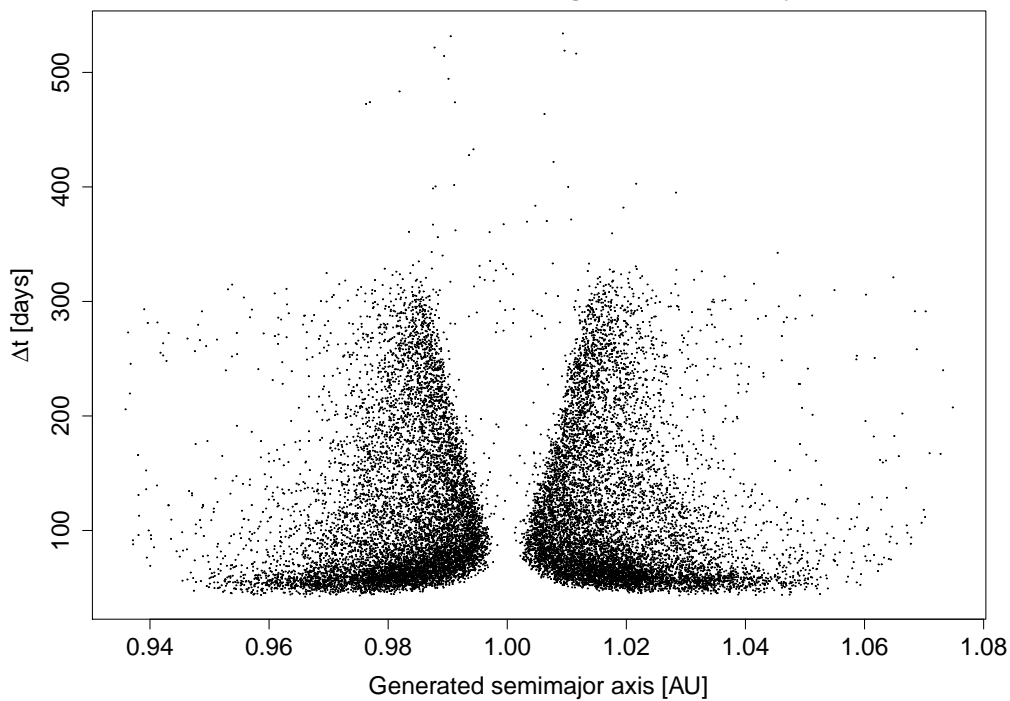}
  \caption{The time it takes for a TP to fulfill the
    temporarily-captured orbiter criteria starting from the generation
    epoch. Most temporarily-captured orbiters are detected if the TPs
    are integrated for more than approximately 500 days.  We
    integrated all the TPs for at least 2,000
    days.}\label{fig:nesc_init_a_dt}
\end{figure}

\newpage
\begin{figure}[H]
  \centering
  \includegraphics[height=\textwidth,angle=-90]{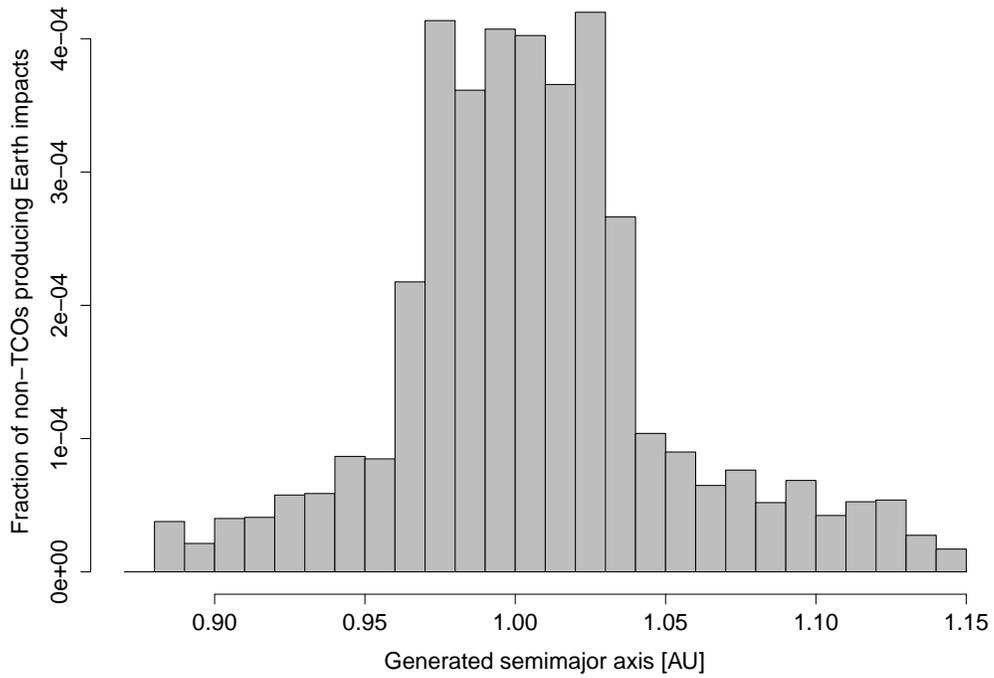}
  \caption{The fraction of TPs colliding with the Earth during a 2,000
    day integration as a function of
    $a_\mathrm{h}$.}\label{fig:collisions_a_histo}
\end{figure}

\newpage
\begin{figure}[H]
  \centering
  \includegraphics[width=\textwidth]{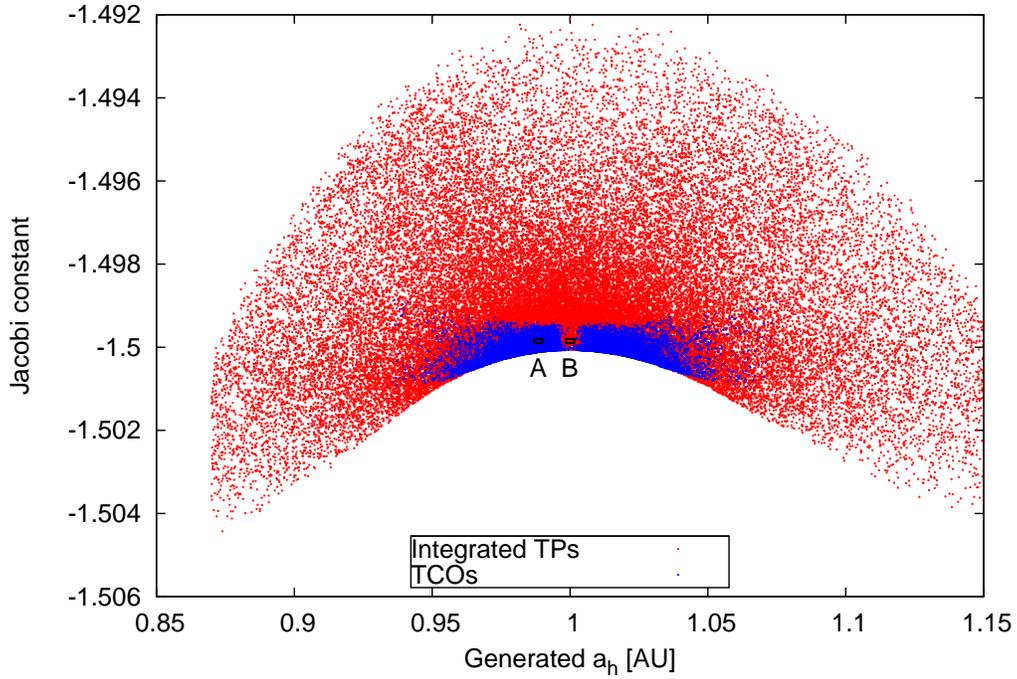}
  \caption{The Jacobi constant, $C$, as a function of $a_\mathrm{h}$
    for representative samples of (red) the integrated TPs and (blue)
    the temporarily-captured orbiters. The black frames labeled 'A'
    and 'B' mark the regions where two random sets of 100 TPs where
    chosen to show that the lack of TCOs with $a_\mathrm{h}\sim1\AU$
    is related to three-body dynamics (see first paragraph in
    Sect.\ \ref{subsec:capture_mechanism} for details). Note that the
    lack of TCOs in 'B' cannot be explained by too-small Jacobi
    constants that would prevent TPs from becoming TCOs in that region
    because the range in $C$ is identical in both
    regions.}\label{fig:aC}
\end{figure}

\newpage
\begin{figure}[H]
  \centering
  \includegraphics[width=\textwidth]{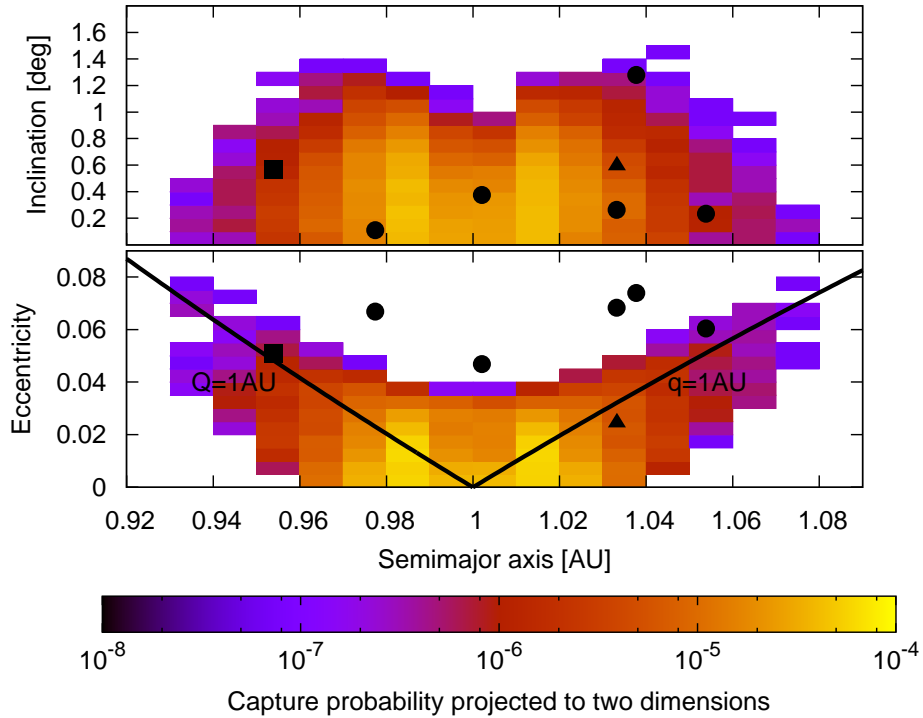}
  \caption{The capture probability in heliocentric
    ($a_\mathrm{h},e_\mathrm{h},i_\mathrm{h}$) space for
    temporarily-captured orbiters (TCOs). The wings of the
    ($a_\mathrm{h},e_\mathrm{h}$) distribution follow the $q=1\AU$ and
    $Q=1\AU$ lines indicating that a TP on an orbit allowing a grazing
    encounter with the Earth's orbit may lead to the TP becoming a TCO
    whereas Earth-crossing orbits in general are not necessarily
    capturable even with small eccentricities. The pre-capture and
    current (post-capture) orbital elements for 2006~RH$_{120}$ are
    indicated with a black square and black triangle,
    respectively. The current orbital elements for other known NEOs
    are represented by black circles. The rightmost object in the
    capturable region is 2007~UN$_{12}$.}\label{fig:ceff_tco_aei}
\end{figure}

\newpage
\begin{figure}[H]
  \centering
  \includegraphics[width=\textwidth]{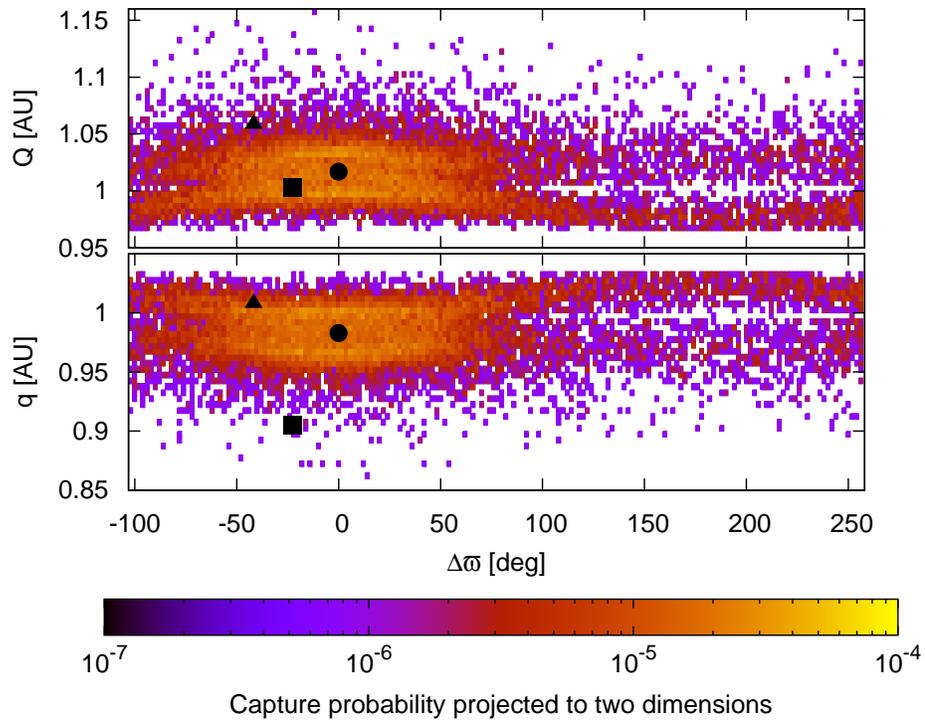}
  \caption{The capture probability in heliocentric
    ($q_\mathrm{h},Q_\mathrm{h},\varpi_\mathrm{h}-\varpi_\oplus$)
    space for temporarily-captured orbiters (TCOs). The pre-capture
    and current (post-capture) orbital elements for 2006~RH$_{120}$
    are indicated with a black square and black triangle,
    respectively. The Earth's orbital elements are marked with a
    filled black circle.}\label{fig:ceff_tco_dlperqQ}
\end{figure}

\newpage
\begin{figure}[H]
  \centering
  \includegraphics[width=\textwidth]{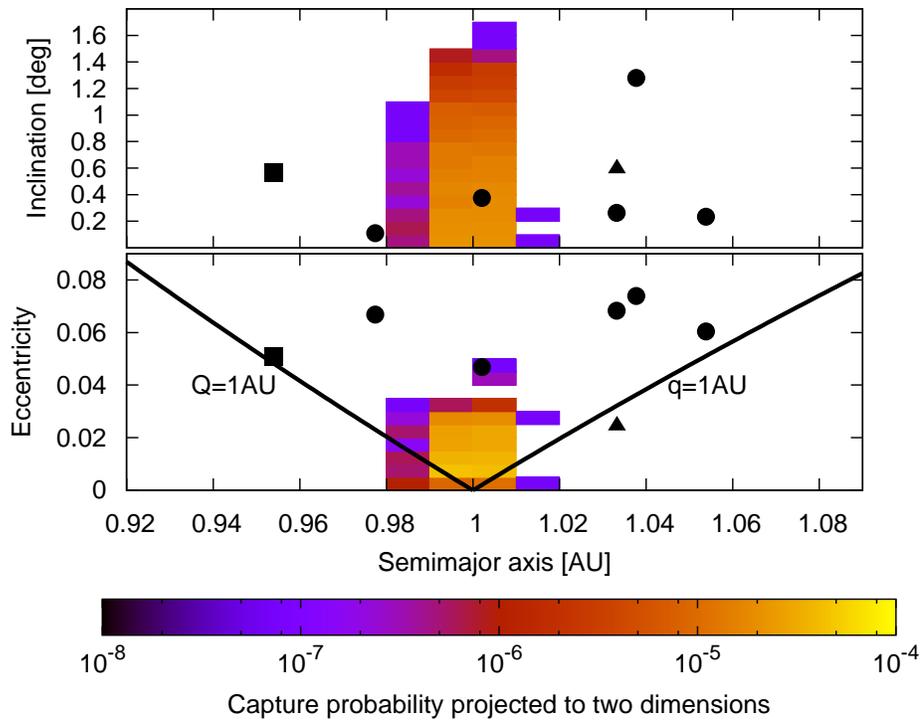}
  \caption{Same as Fig.\ \ref{fig:ceff_tco_aei} but for the
    barycentric dynamical model that does not include the Moon. We
    note that both the pre-capture orbit and the current
    (post-capture) orbit for 2006 RH$_{120}$ are outside of the region
    harboring capturable orbits.}\label{fig:ceff_tco_aei_bary}
\end{figure}

\newpage
\begin{figure}[H]
  \centering
  \includegraphics[width=0.8\textwidth]{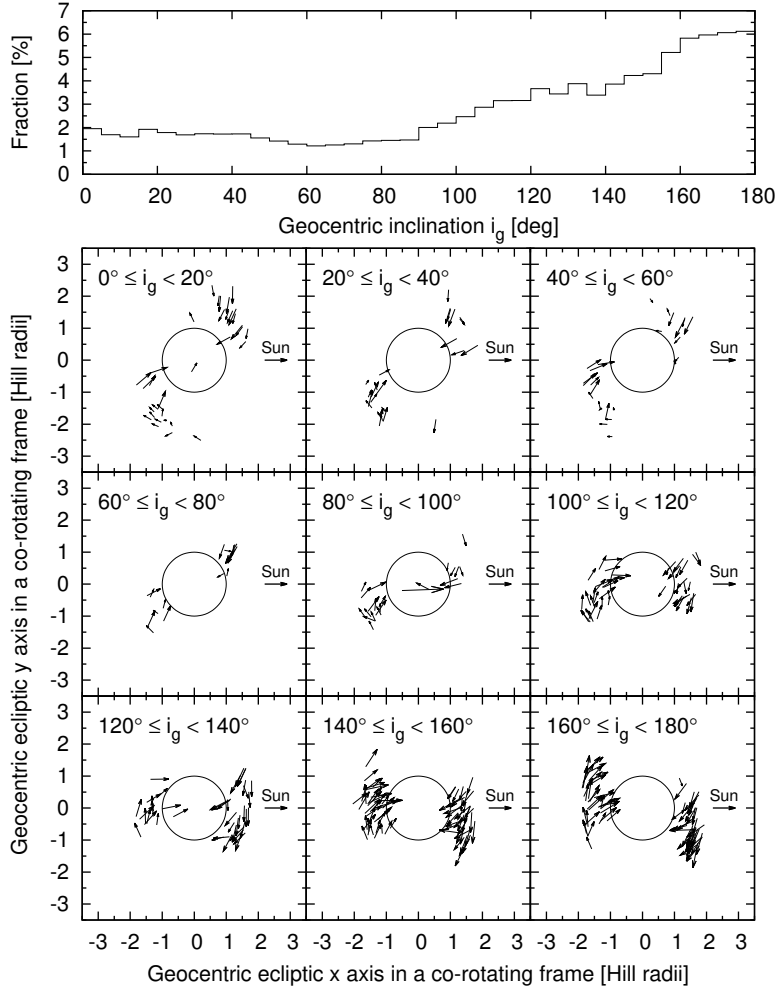}
  \caption{(Top) Fractional distribution of the TCO geocentric
    inclination distribution.  Most TCOs are in retrograde orbits
    typical of irregular satellites.  (Bottom) Positions and
    instantaneous velocity vectors of temporarily-captured orbiters at
    time of capture projected onto the ecliptic plane in a co-rotating
    coordinate system as a function of inclination. For correct
    interpretation, note that both the prograde ($i_\mathrm{g} <
    90\deg$) and retrograde ($i_\mathrm{g} > 90\deg$) TCOs typically
    move in a retrograde fashion in the co-rotating system: the
    prograde ones with an average angular rate of
    $-0.43\deg\days^{-1}$ and the retrograde ones with an average
    angular rate of
    $-1.6\deg\days^{-1}$.}\label{fig:rotcoord_at_capture}
\end{figure}

\newpage
\begin{figure}[H]
  \centering
  \includegraphics[width=0.75\textwidth]{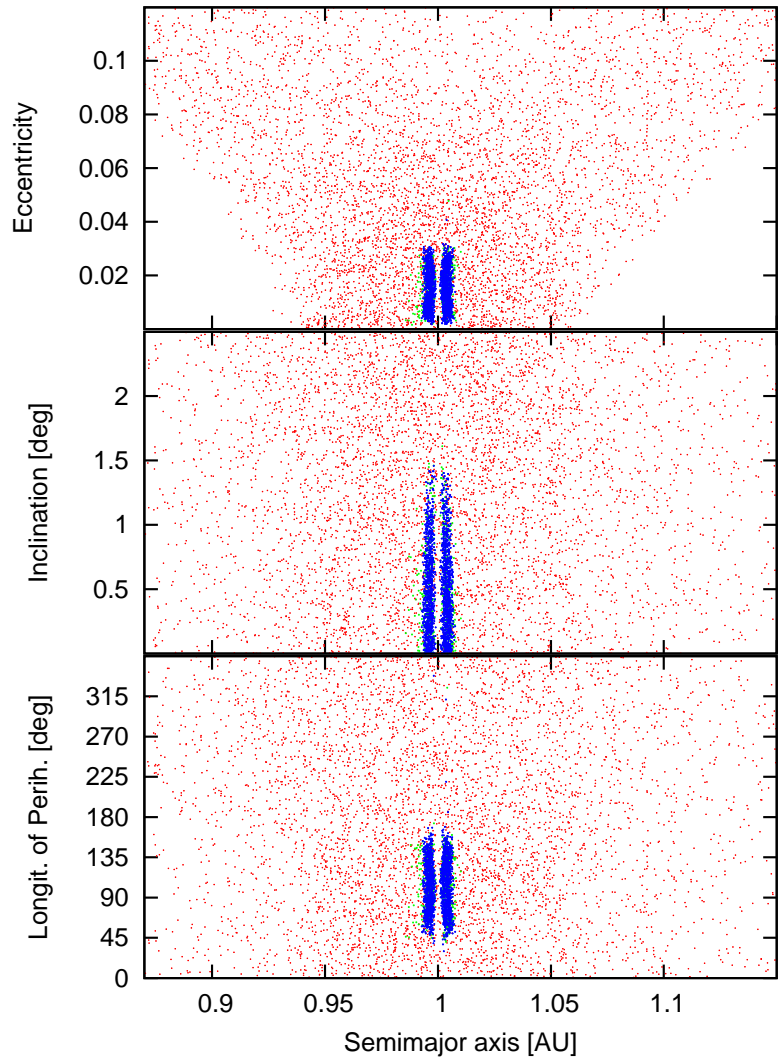}
  \caption{The same as Fig.\ \ref{fig:nesc_init_elements} but for the
    barycentric model where the Moon has been
    omitted.}\label{fig:nesc_init_elements_bary}
\end{figure}

\newpage
\begin{figure}[H]
  \centering
  \includegraphics[height=\textwidth,angle=-90]{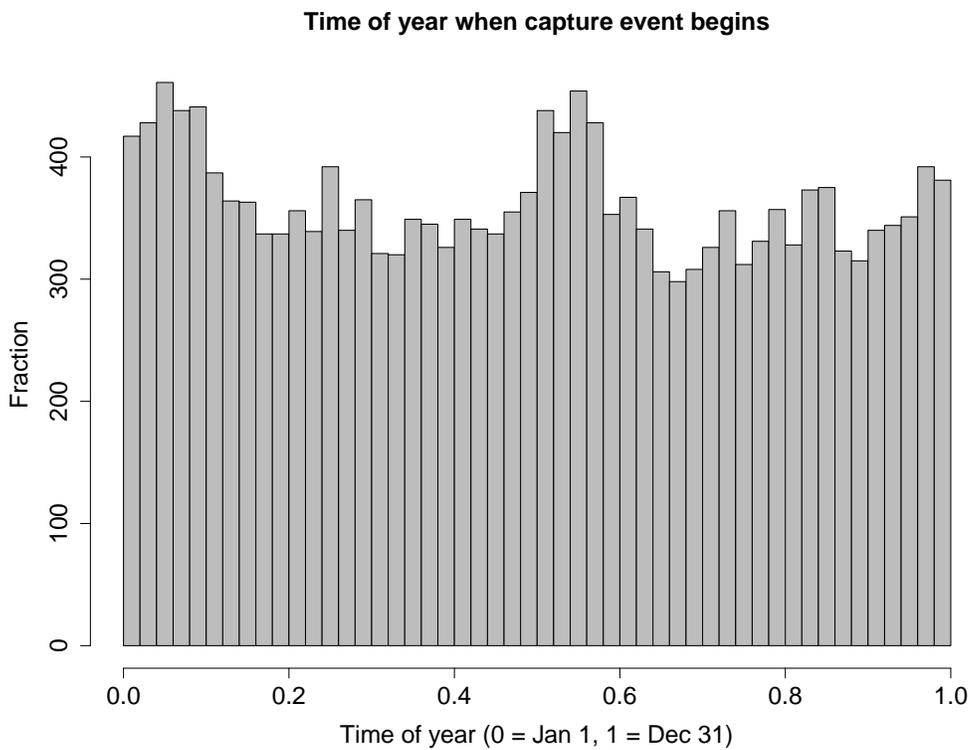}
  \caption{The time of year a temporarily-captured orbiter is
    captured. The distribution of generation epochs was uniform but
    the time-of-capture distribution has maxima in January and
    July.}\label{fig:tco_time_of_capture}
\end{figure}

\newpage
\begin{figure}[H]
  \centering
  \includegraphics[width=\textwidth]{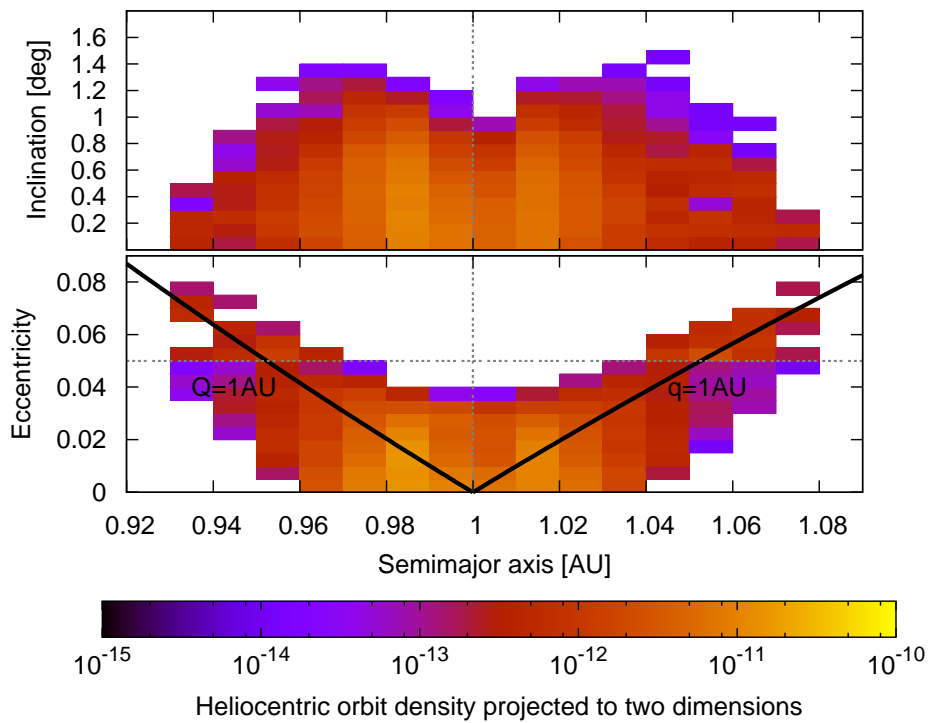}
  \caption{The heliocentric orbit-density distribution for TCOs at the
    generation epoch, i.e., the product of the capture-probability
    distribution (see Fig.\ \ref{fig:ceff_tco_aei}) and the debiased
    NEO orbit-density distribution \protect \cite{bot2002a}. The
    dotted gray lines indicate the resolution of the NEO orbit-density
    distribution.}\label{fig:tco_hodd}
\end{figure}

\newpage
\begin{figure}[H]
  \centering
  \includegraphics[width=0.7\textwidth,angle=-90]{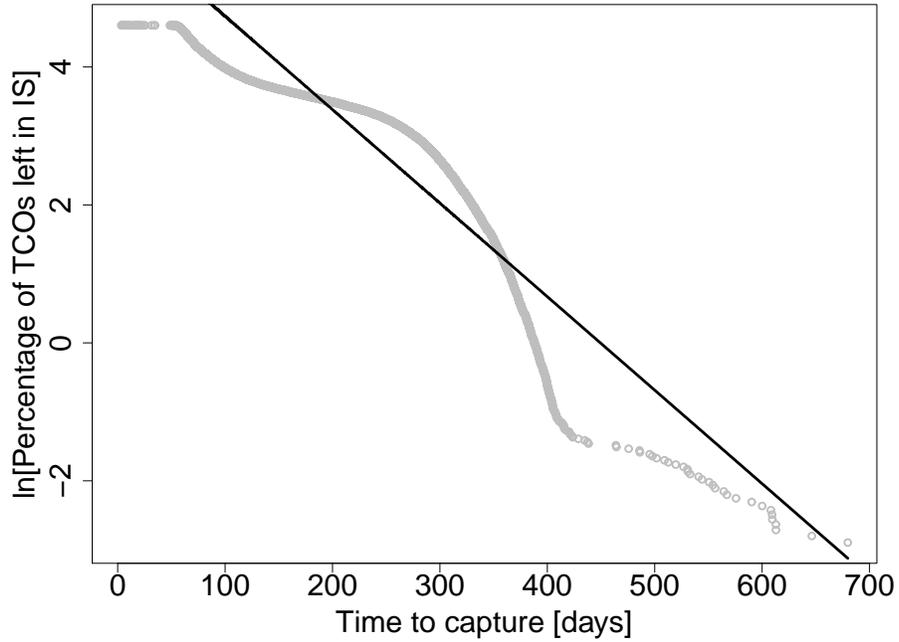}
  \caption{Decay from the outer edge of the intermediate source
    population (the generated NEOs propagated out to 5 Earth Hill
    radii from the Earth) into the temporarily-captured orbiter
    population. The black line represents a number-weighted fit for
    the slope (see Sect.\ \ref{subsec:steady-state} for details). Note
    that most of the weight is approximately in the interval from 55
    to 85 days.}\label{fig:decay}
\end{figure}

\newpage
\begin{figure}[H]
  \centering
  \includegraphics[width=\textwidth]{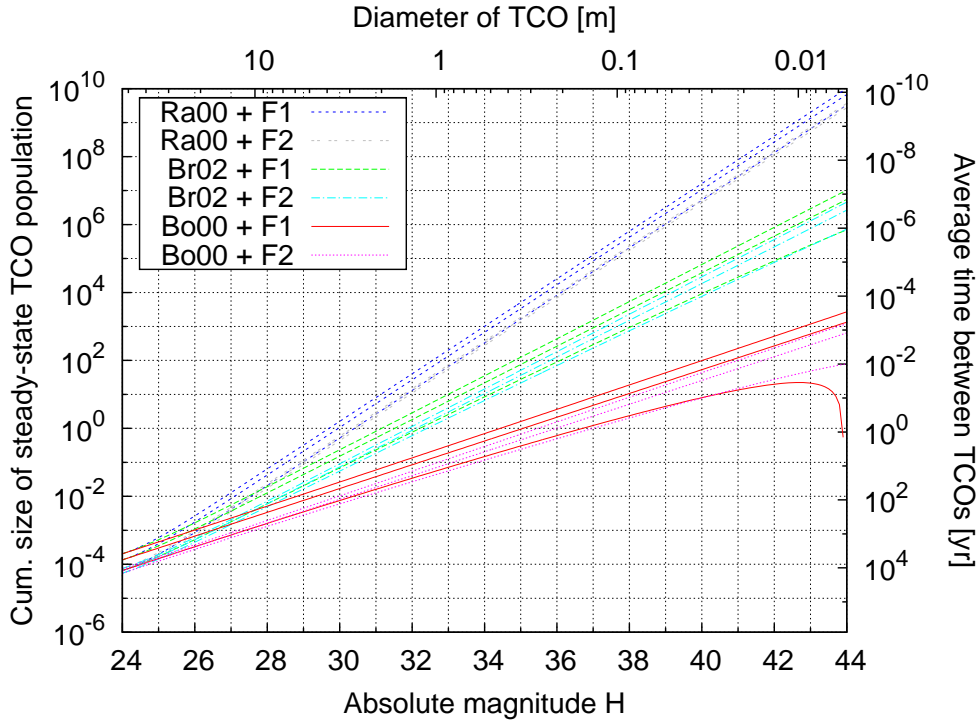}
  \caption{The cumulative steady-state SFD, $N(<H)$, for
    temporarily-captured orbiters using three different NEO SFDs and
    two different methods of calculating $F_\mathrm{TCO}$ (F1 refers
    to the method used by \protect \citeasnoun{mora2002a} and F2 to
    the alternative method presented in this work). In the text we
    argue that F2 and the \protect \citeasnoun{bro2002a} NEO SFD
    provide the most robust estimate which means that the maximum size
    at which at least one object is captured at any given time is
    $H\sim32$ (or a diameter of approximately 1\,m), and that the
    frequency of temporarily-captured orbiters with $H\sim30$ is about
    one every decade. The uncertainty envelopes correspond to the
    1-$\sigma$ uncertainties for the size of the steady-state
    population that incorporate uncertainty estimates for all other
    factors but the NEO orbit distribution by \protect
    \citeasnoun{bot2002a}, the capture-efficiency distribution, and
    the slope for the NEO SFD by \protect \citeasnoun{rab2000a}. Note
    that the envelopes can thus not be directly used to estimate the
    uncertainty of the frequency of temporarily-captured orbiters on
    the RHS axis. The conversion from $H$ magnitude to diameter
    assumes a geometric albedo of 0.15.}\label{fig:tco_steady-state}
\end{figure}

\newpage
\begin{figure}[H]
  \centering
  \includegraphics[width=\textwidth]{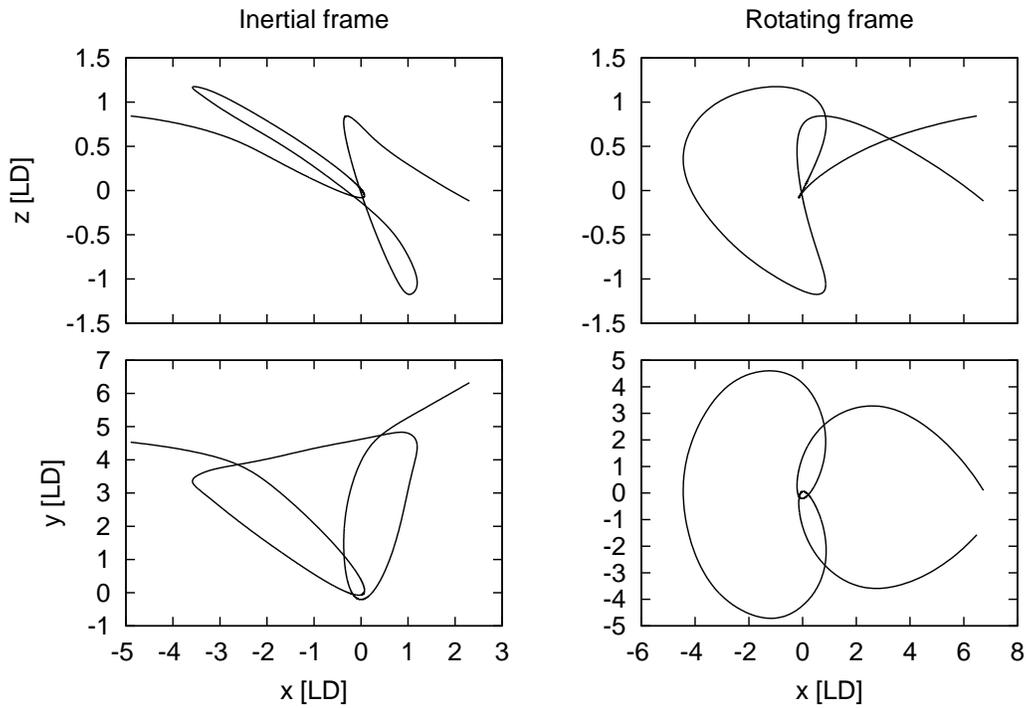}
  \caption{The trajectory of a temporarily-captured orbiter in (left
    column) geocentric, inertial Cartesian coordinates and (right
    column) geocentric, co-rotating Cartesian coordinates. This
    particular TP was chosen because of its close-to-average
    characteristics: during its 280-day capture it makes 2.94
    retrograde revolutions around the Earth as measured in a
    co-rotating frame. The distance scale is given in lunar distances
    (LD) equaling approximately
    $0.00257\AU$.}\label{fig:typical_tco_xyz}
\end{figure}

\newpage
\begin{figure}[H]
  \centering
  \includegraphics[width=\textheight,angle=-90]{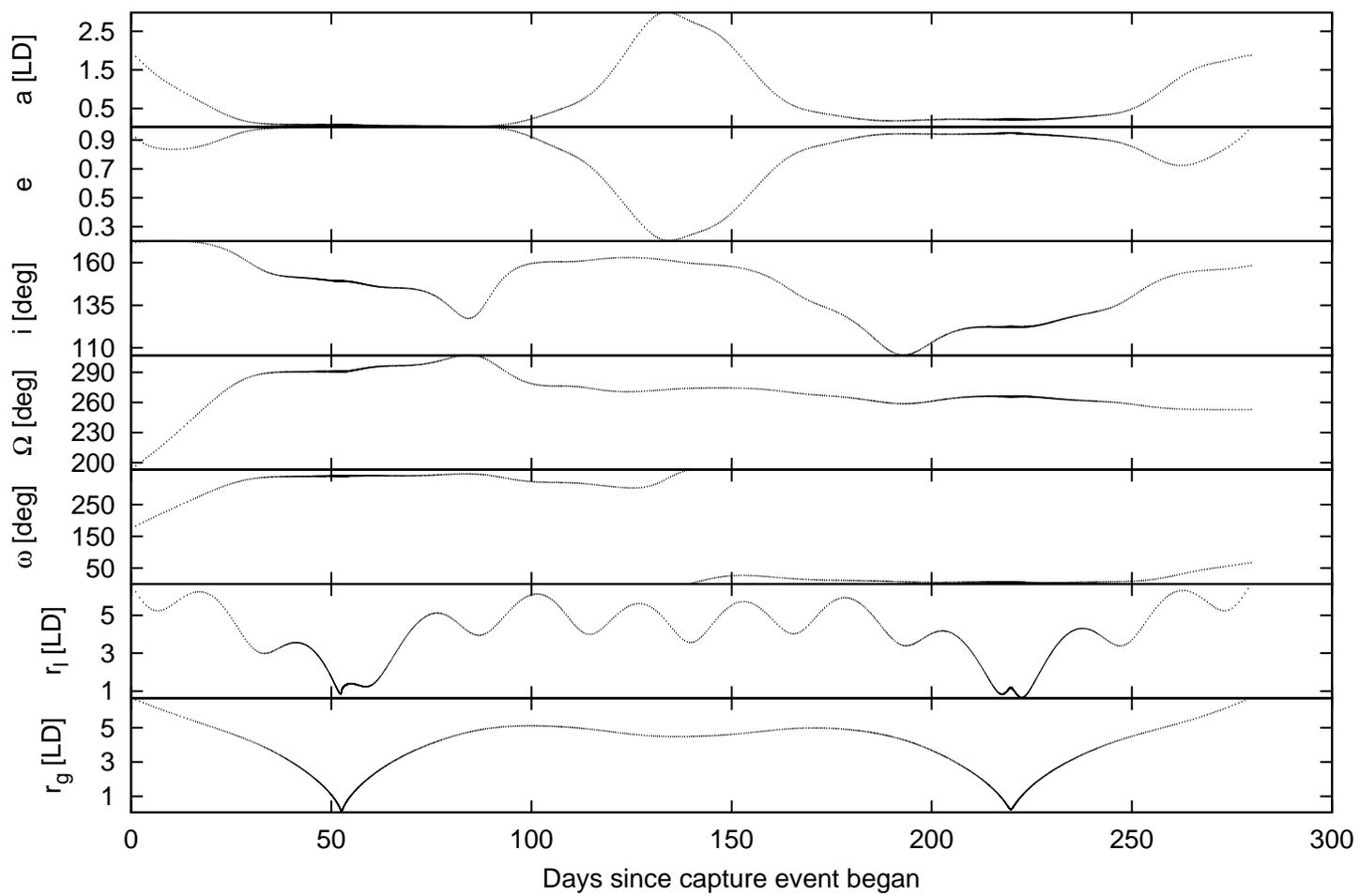}
  \caption{The evolution of the geocentric Keplerian elements and the
    lunacentric and geocentric distances for the same object as in
    Fig.\ \ref{fig:typical_tco_xyz}. The distance scale is in lunar
    distances (LD) equaling $\sim
    0.00257\AU$.}\label{fig:typical_tco_kep}
\end{figure}

\newpage
\begin{figure}[H]
  \centering
  \includegraphics[width=\textwidth]{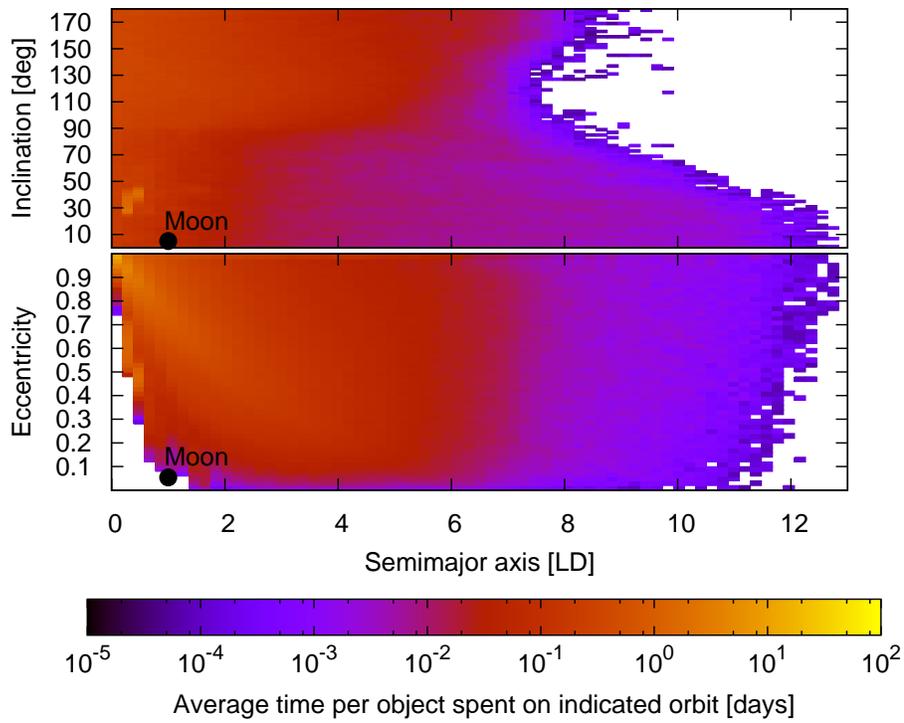}
  \caption{The residence time as a function of geocentric
    ($a_\mathrm{g},e_\mathrm{g},i_\mathrm{g}$) for
    temporarily-captured orbiters.}\label{fig:tco_residence_aei}
\end{figure}

\newpage
\begin{figure}[H]
  \centering
  \includegraphics[width=\textwidth]{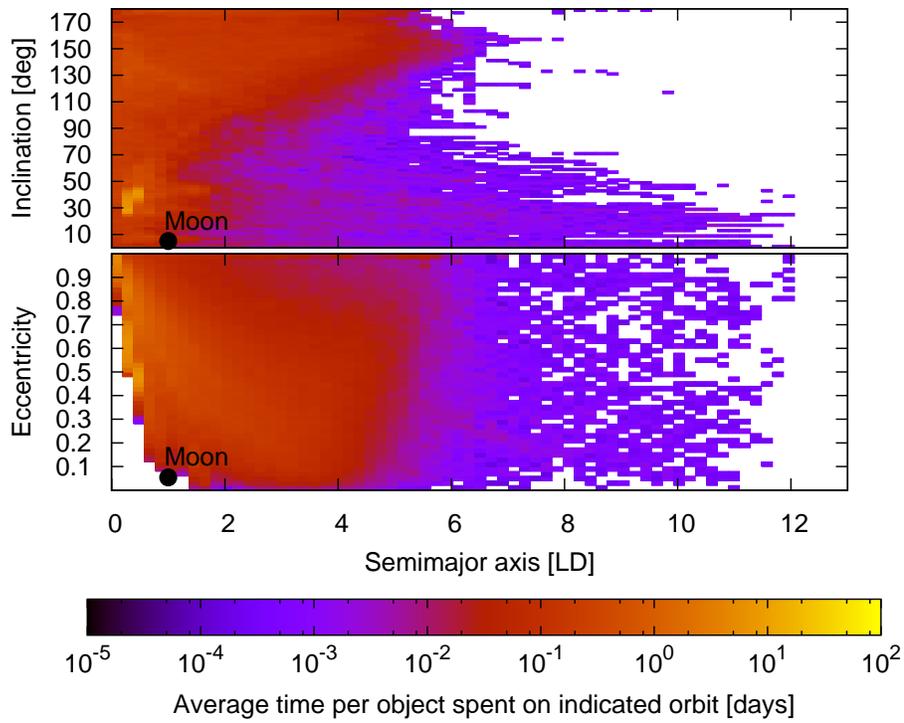}
  \caption{The residence time as a function of geocentric
    ($a_\mathrm{g},e_\mathrm{g},i_\mathrm{g}$) for the subset of
    temporarily-captured orbiters that make at least five revolutions
    around the Earth.}\label{fig:tco_residence_gt5rev_aei}
\end{figure}

\newpage
\begin{figure}[H]
  \centering
  \includegraphics[width=\textwidth]{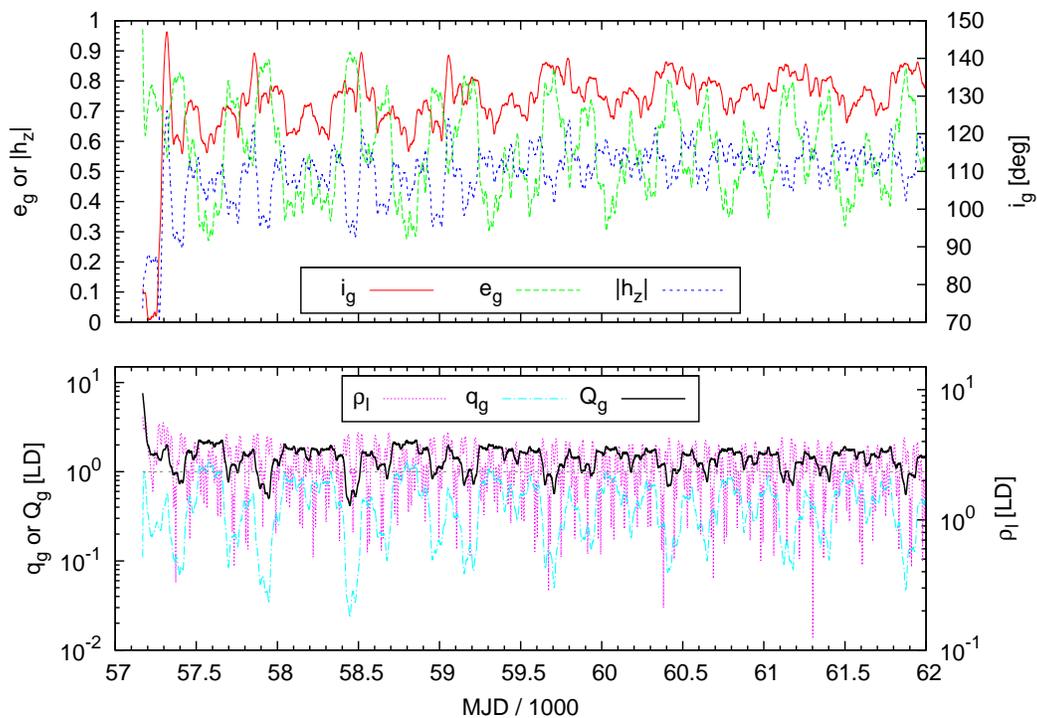}
  \caption{Geocentric orbital-element evolution for a
    temporarily-captured orbiter that is strongly affected by the
    Kozai resonance (inclination $i_\mathrm{g}$, eccentricity
    $e_\mathrm{g}$, the normal component of the angular momentum
    $|h_\mathrm{z}|$, lunacentric distance $\rho_\mathrm{l}$, perigee
    $q_\mathrm{g}$, and the apogee $Q_\mathrm{g}$). The figure
    represents the first 13 years of a 35-year-long capture. The
    horizontal dashed line in the bottom figure represents one lunar
    distance.}\label{fig:xyGf_qQplot_mjdlt62000}
\end{figure}

\newpage
\begin{figure}[H]
  \centering
  \includegraphics[width=\textwidth]{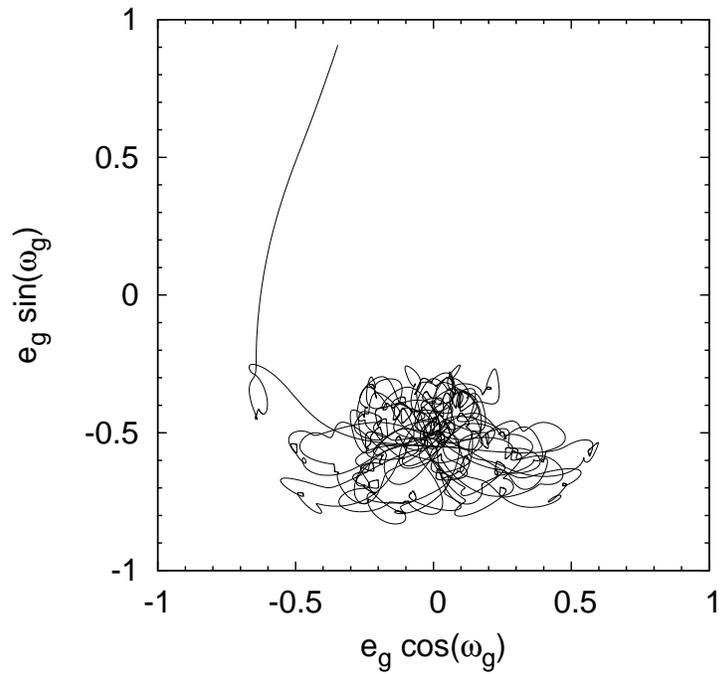}
  \caption{The polar plot for the object in
    Fig.\ \ref{fig:xyGf_qQplot_mjdlt62000}. The geocentric orbit is
    strongly perturbed by the Sun. The argument of pericenter is
    librating around $\omega_\mathrm{g}=270\deg$ which, combined with the
    $e_\mathrm{g}$-$i_\mathrm{g}$ oscillation shown in
    Fig.\ \ref{fig:xyGf_qQplot_mjdlt62000}, reveals that this TP is
    strongly affected by the Kozai
    resonance.}\label{fig:xyGf_polarplot}
\end{figure}

\newpage
\begin{figure}[H]
  \centering
  \includegraphics[height=\textwidth,angle=-90]{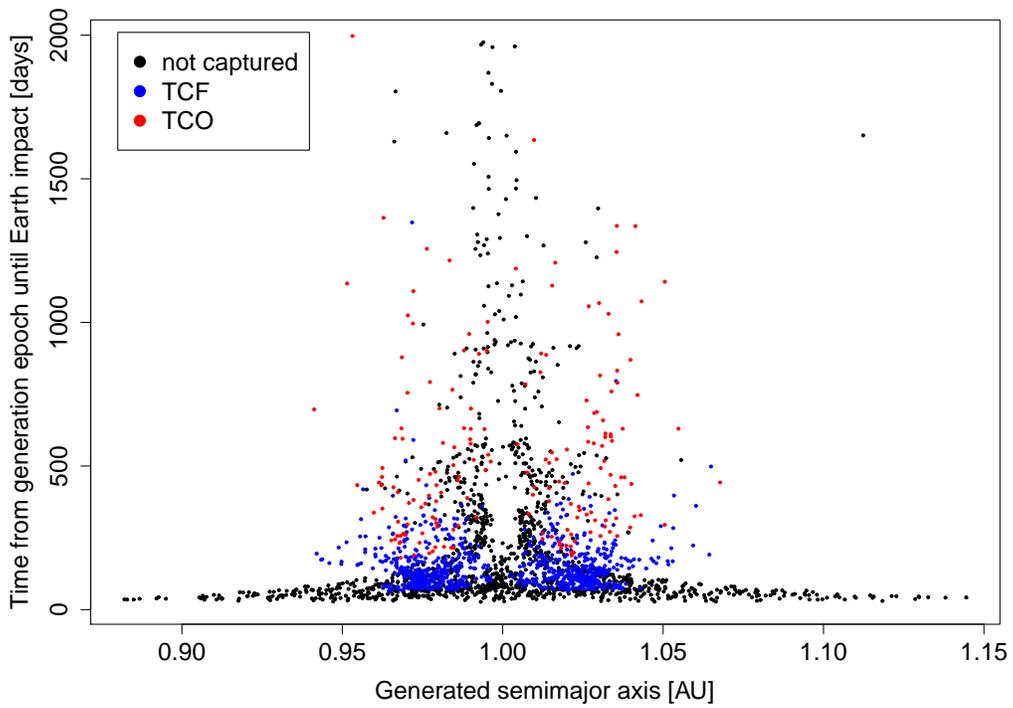}
  \caption{The time from generation epoch to Earth impact as a
    function of the generated semimajor axis for (red)
    temporarily-captured orbiters, (blue) temporarily-captured flybys,
    and (black) uncaptured objects.}\label{fig:collisions_a_dt}
\end{figure}

\newpage
\begin{figure}[H]
  \centering
  \includegraphics[height=\textwidth,angle=-90]{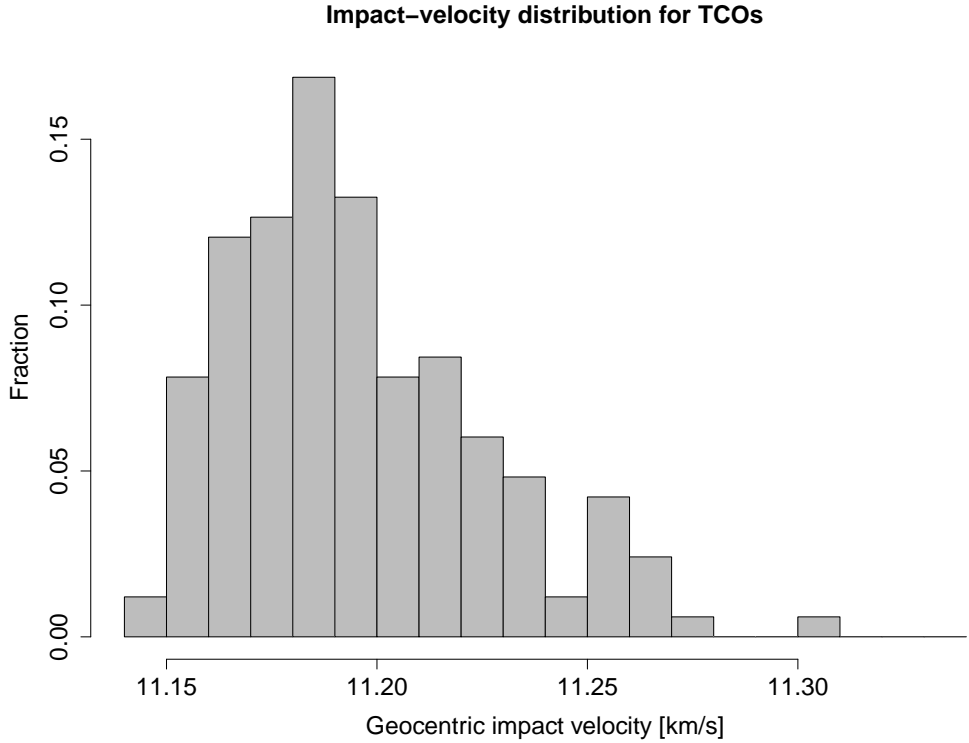}
  \caption{The Earth-impact speed distribution for
    temporarily-captured orbiters. The Earth's escape speed is
    $11.2\km\second^{-1}$. The Earth-impact speed estimates come from
    orbital integrations for which the requirement on the relative
    accuracy was set to $10^{-10}$.}\label{fig:collisions_vel_histo}
\end{figure}

\newpage
\begin{figure}[H]
  \centering
  \includegraphics[width=\textwidth]{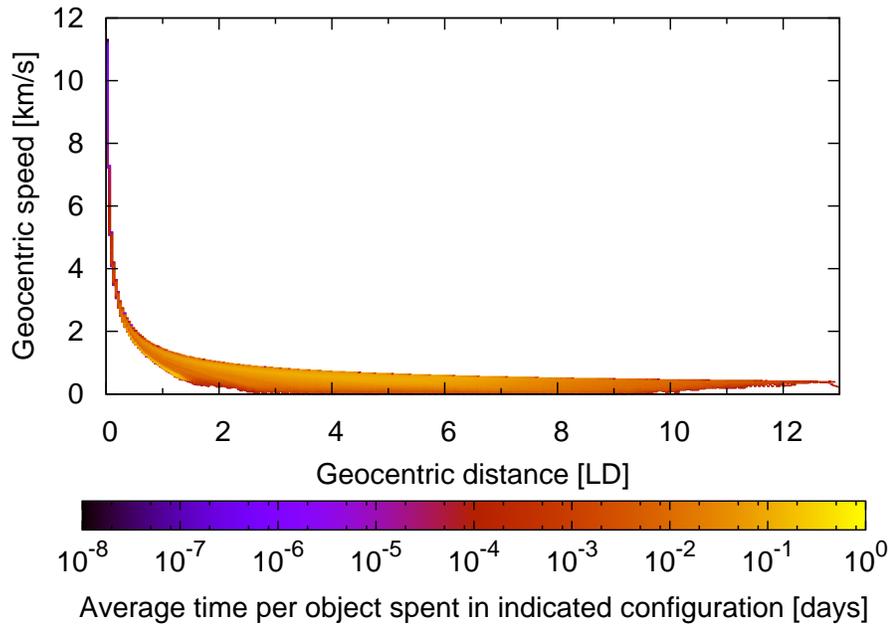}
  \caption{The TCO residence time as a function of geocentric
    ($r_\mathrm{g},v_\mathrm{g}$).  The implication is that TCOs will
    be difficult to detect from the ground because they are moving
    fast when they are closest to the Earth and thus at their
    brightest.}\label{fig:tco_residence_rv}
\end{figure}

\newpage
\begin{figure}[H]
  \centering
  \includegraphics[height=\textwidth,angle=-90]{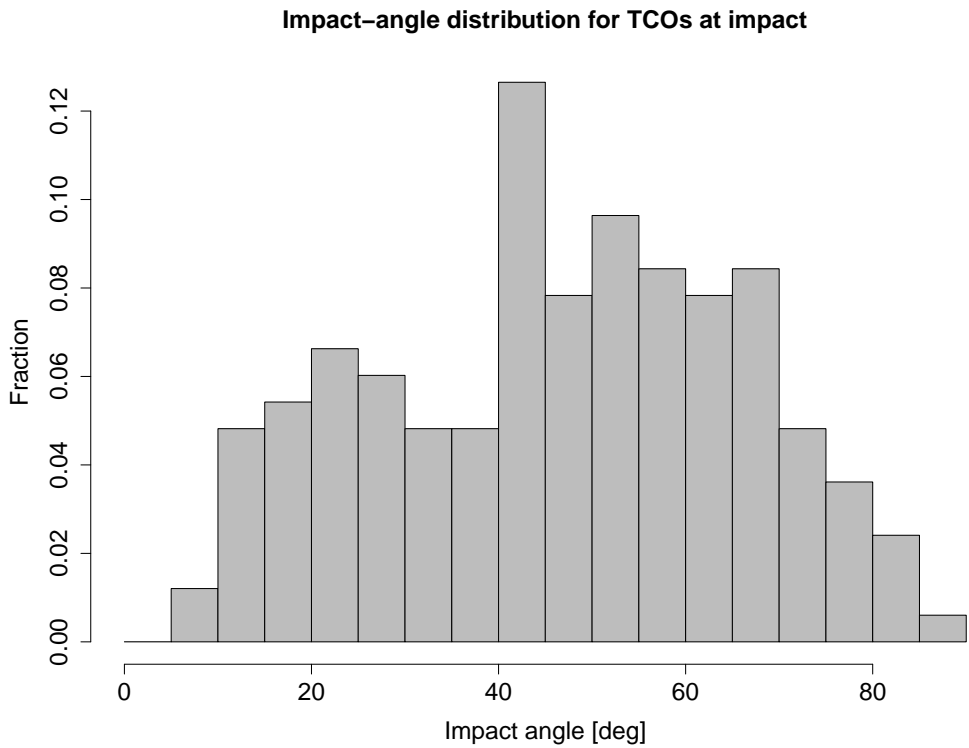}
  \caption{Impact-angle distribution for Earth-impacting
    temporarily-captured orbiters at the time of hitting the Earth's
    surface.}\label{fig:collisions_impact_angle_histo}
\end{figure}

\newpage
\begin{figure}[H]
  \centering
  \includegraphics[width=\textwidth]{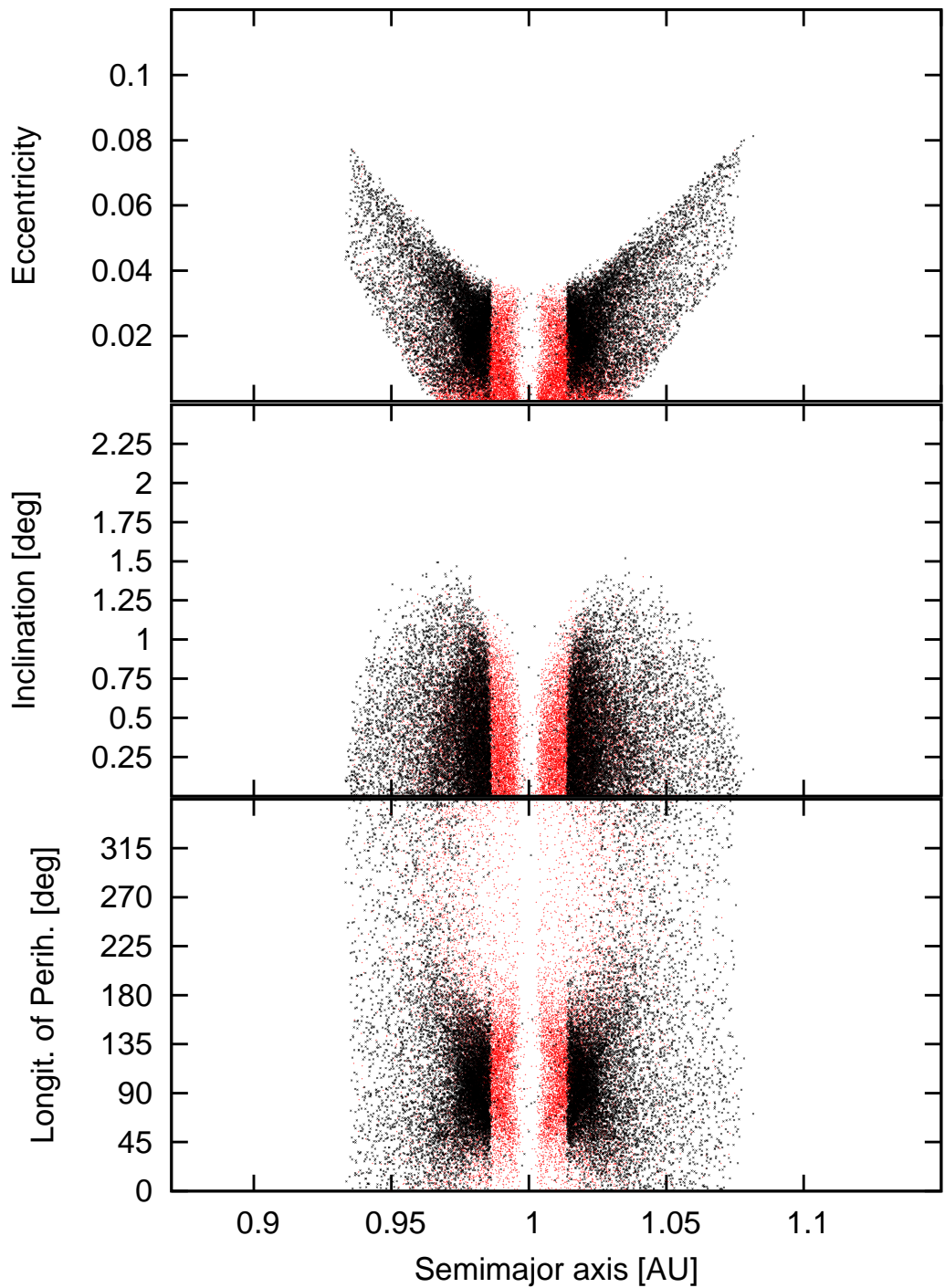}
  \caption{(black) Orbital elements for temporarily-captured orbiters one year
    after their escape from the EMS and (red) the same objects'
    elements prior to capture.}\label{fig:TCO_postcapture_elements}
\end{figure}

\newpage
\begin{figure}[H]
  \centering
  \includegraphics[width=\textwidth]{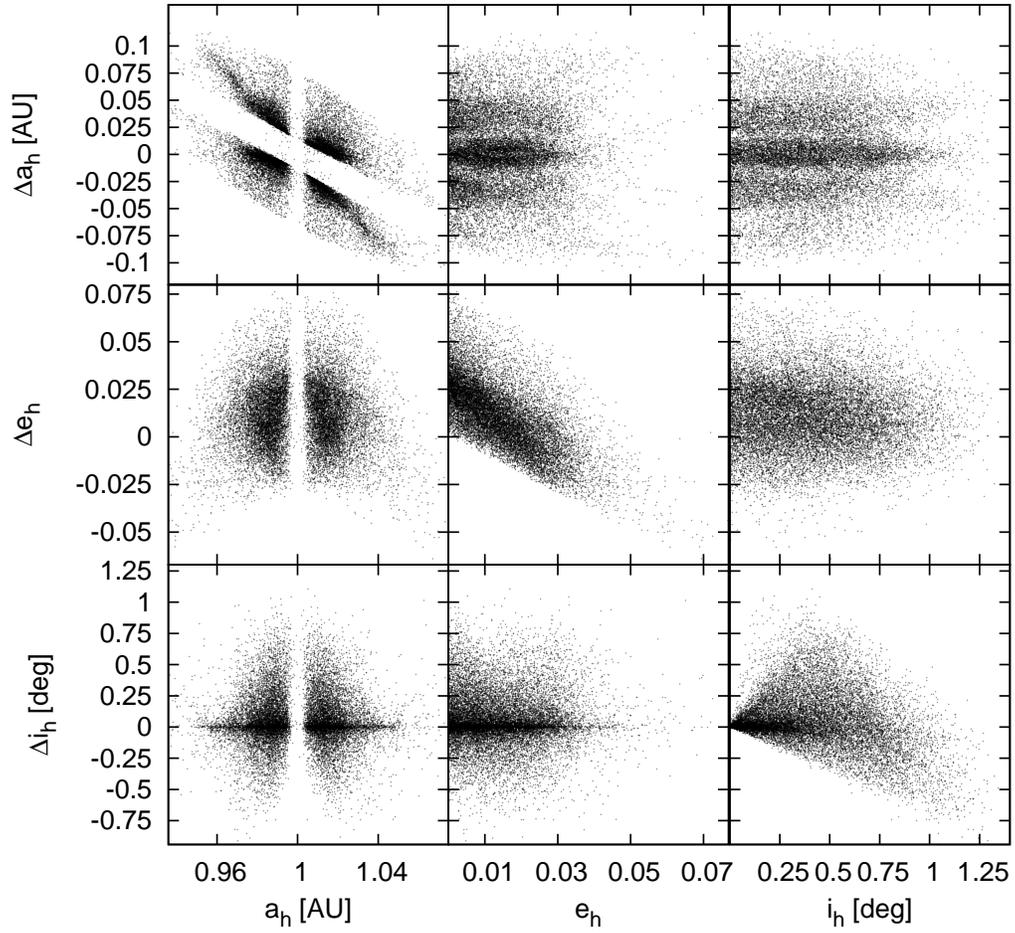}
  \caption{The change in orbital elements for temporarily-captured
    orbiters between the generation epoch (prior to capture) and the
    epoch one year after the escape from the
    EMS.}\label{fig:TCO_postcapture_delements}
\end{figure}

%
%
%
%

\newpage 
\appendix

\section{A method for computing a 6D orbit given
  ($a$,$e$,$i$) and a position vector $\vect{r}$}
\label{appsec:initialconditions}

In this work the generation of the `intermediate source population'
requires that we generate NEOs with a known ($a$,$e$,$i$) distribution
within a shell from 4-5 Hill radii of the Earth.  Here we present an
analytical technique for generating these objects rather than the
`brute-force' technique employed above.

Given ($a$,$e$,$i$) and the position vector from the central body to
the object of interest at epoch $t_0$ $\vect{r}(t_0)=(x,y,z)_{t_0}$ to
calculate the corresponding 6D orbit we need to solve for either
$\dot{\vect{r}}(t_0)$ or ($\Omega$,$\omega$,$M_0$), where $\Omega$ is
the longitude of ascending node, $\omega$ is the argument of
pericenter, and $M_0$ is the mean anomaly at the chosen epoch. In what
follows we solve for the latter as this choice results in a more
elegant method.

An object's cartesian coordinates in the orbital plane at the epoch
time, $\vect{r}_p(t_0)=(x_p,y_p,z_p)_{t_0}$, are given by
\begin{equation}\label{eq:orbplane}
  \begin{split}
    x_p & =  a (\cos{E} - e)\,, \\
    y_p & =  a \sqrt{1- e^2}\sin{E} = a \sqrt{(1- e^2)(1-\cos^2{E})}\,, \\
    z_p & =  0\,,
  \end{split}
\end{equation}
where $E$ is the eccentric anomaly and
\begin{equation}\label{eq:ea}
  \cos{E} = \frac{1 - \frac{|\vect{r}|}{a}}{e}
\end{equation} 
Combining Eq.\ \ref{eq:ea} and Kepler's equation yields two possible
values for $M_0$. 

The coordinates in the orbital plane $\vect{r}_p(t_0)$ are connected
to the coordinates in the ecliptic plane $\vect{r}(t_0)$ through the
$3\times3$ rotation matrix $R$
\begin{equation}\label{eq:rotation}
\vect{r}(t_0) = R\vect{r}_p(t_0)\,,
\end{equation} 
with
\begin{equation}\label{eq:rotmat}
  \begin{split}
    R_{11} & =  \cos{\Omega}\cos{\omega} - \sin{\Omega}\sin{\omega}\cos{i} \\
    R_{12} & =  -\cos{\Omega}\sin{\omega} - \sin{\Omega}\cos{\omega}\cos{i} \\
    R_{13} & =  \sin{\Omega}\sin{i} \\
    R_{21} & =  \sin{\Omega}\cos{\omega} + \cos{\Omega}\sin{\omega}\cos{i} \\
    R_{22} & =  -\sin{\Omega}\sin{\omega} + \cos{\Omega}\cos{\omega}\cos{i} \\
    R_{23} & =  -\cos{\Omega}\sin{i} \\
    R_{31} & =  \sin{\omega}\sin{i} \\
    R_{32} & =  \cos{\omega}\sin{i} \\
    R_{33} & =  \cos{i}\,.
  \end{split}
\end{equation}

Combining Eqs.\ \ref{eq:orbplane}--\ref{eq:rotmat} we derive the
following separate equations for $\cos{\Omega}$ and $\cos{\omega}$
\begin{align}
  & A_\Omega\sqrt{1-\cos^2{\Omega}} - \cos{\Omega} + B_\Omega = 0\,, \label{eq:cosnode} \\
  & A_\omega \sqrt{1-\cos^2{\omega}} + B_\omega \cos{\omega} - C_\omega = 0\,,  \label{eq:cosperi}
\end{align}
where
\begin{align*}
  A_\Omega &= \frac{x}{y}\,, \\
  B_\Omega &= \frac{z}{y\tan{i}}\,, \\
  A_\omega &= a \left(\frac{1 - \frac{|\vect{r}|}{a}}{e} - e\right)\,, \\
  B_\omega &= a \sqrt{(1- e^2)\left( 1 - \left(\frac{1 - \frac{|\vect{r}|}{a}}{e}\right)^2\right)}\,, \\
  C_\omega &= \frac{z}{\sin{i}}\,.
\end{align*}

$B_\Omega$ and $C_\omega$ have singularities at $i=0\deg$ that
correspond to $\Omega$ and $\omega$ being undefined for
zero-inclination orbits. Similarly, $A_\omega$ and $B_\omega$ have
singularities at $e=0$ that correspond to $\omega$ being undefined
for circular orbits. $B_\Omega$ also has a singularity at $y=0$. 

The solutions to Eqs.\ \ref{eq:cosnode} and \ref{eq:cosperi} are
simply obtained from the quadratic equation:
\begin{align}
  & \cos{\Omega} = \frac{B_\Omega \pm A_\Omega \sqrt{A_\Omega^2 - B_\Omega^2 + 1}}{A_\Omega^2 + 1}\,, \label{eq:cosnodesol} \\
  & \cos{\omega} = \frac{B_\omega C_\omega \pm A_\omega \sqrt{A_\omega^2 + B_\omega^2 - C_\omega^2}}{A_\omega^2 + B_\omega^2}\,. \label{eq:cosperisol} 
\end{align}
The right-hand sides of Eqs.\ \ref{eq:cosnodesol} and
\ref{eq:cosperisol} have a non-zero imaginary part if i) the
pericenter distance $q=a(1-e)$ is larger than $|\vect{r}(t_0)|$, ii)
the apocenter distance $Q = a(1+e)$ is smaller than $|\vect{r}(t_0)|$,
and/or iii) the $z$-component of the position vector is larger than
$|\vect{r}(t_0)|\sin i$. Since solutions with non-zero imaginary parts
are non-physical, we choose only the real valued solutions and
determine which permutations of
($a$,$e$,$i$,$\Omega_j$,$\omega_k$,$M_{0,m}$), where $j=1,2,3,4$,
$k=1,2,3,4$, $m=1,2$, result in
\begin{equation}
|\vect{r}(t_0)-\mathrm{pos}(a,e,i,\Omega,\omega,M_0,t_0)| < \epsilon\,,
\end{equation}
where the operator $\mathrm{pos}$ converts Keplerian elements to a
Cartesian position, and $\epsilon$ is a small positive quantity. For a
given combination of $a$, $e$, $i$ and $\vect{r}(t_0)$ we get either
zero or four orbits \citeaffixed{jed1996a}{cf.\ }.

\end{document}